\documentclass{article}

\usepackage[utf8]{inputenc}
\usepackage{listings}
\usepackage{color}
\usepackage{fullpage}
\usepackage{color}
\usepackage{amssymb,graphicx}
\usepackage{float}
\usepackage{verbatim}
\floatstyle{plain}
\restylefloat{figure}
\usepackage{caption}
\usepackage{subcaption}
\usepackage{tabu}
\usepackage{graphicx}
\usepackage{hyperref}
\usepackage{amsthm}
\usepackage{amsmath}

\usepackage{algorithmicx}
\usepackage{algorithm}
\usepackage[noend]{algpseudocode}
\usepackage{multicol}
\algnewcommand\algorithmicforeach{\textbf{for each}}
\algdef{S}[FOR]{ForEach}[1]{\algorithmicforeach\ #1\ \algorithmicdo}
\usepackage{tikz}
\usepackage{makecell}
\usepackage{comment}
\usepackage{mathtools}
\usepackage{etoolbox}
\usepackage{url}
\makeatletter
\newcommand\footnoteref[1]{\protected@xdef\@thefnmark{\ref{#1}}\@footnotemark}
\makeatother
\theoremstyle{definition}

\theoremstyle{definition}

\theoremstyle{definition}
\newtheorem{definition}{Definition}[subsection]

\date{}
\title{Accelerating Big-Data Sorting Through Programmable Switches}

\author{Yamit Barshatz-Schneor \ \ \ \ \ \ Roy Friedman\\
	Computer Science Department\\
	Technion}

\begin{document}
\maketitle

\begin{abstract}

Sorting is a fundamental and well studied problem that has been studied extensively. 
Sorting plays an important role in the area of databases, as many queries can be served much faster if the relations are first sorted.
One of the most popular sorting algorithm in databases is merge sort.

In modern data-centers, data is stored in storage servers, while processing takes place in compute servers. 
Hence, in order to compute queries on the data, it must travel through the network from the storage servers to the compute servers. 
This creates a potential for utilizing programmable switches to perform partial sorting in order to accelerate the sorting process at the server side. 
This is possible because, as mentioned above, data packets pass through the switch in any case on their way to the server.
Alas, programmable switches offer a very restricted and non-intuitive programming model, which is why realizing this is not-trivial.

We devised a novel partial sorting algorithm that fits the programming model and restrictions of programmable switches and can expedite merge sort at the server.
We also utilize built-in parallelism in the switch to divide the data into sequential ranges. 
Thus, the server needs to sort each range separately and then concatenate them to one sorted stream. 
This way, the server needs to sort smaller sections and each of these sections is already partially sorted. 
Hence, the server does less work, and the access pattern becomes more virtual-memory friendly.

We evaluated the performance improvements obtained when utilizing our partial sorting algorithm over several data stream compositions with various switch configurations. 
Our study exhibits an improvement of 20\%-75\% in the sorting run-time when using our approach compared to plain sorting on the original~stream. 
\end{abstract}
	
\section{Introduction}
\label{sec:intro}


Programmable switches, as their name suggests, can be dynamically programmed to execute user defined functionality.
This capability is supported by major switch vendors~\cite{tofino, broadcom, Cavium}. 
In particular, a programmer can define methods for collecting on-the-fly information regarding packets and then analyze the traffic and deduce network statistics and telemetry, perform network load balancing and detect attacks, thereby improving network performance in real time.

Recent work has demonstrated how programmable switches can be used to accelerate SQL queries in a data-center setting~\cite{tirmazi2020cheetah}.
In data-centers, data is typically stored on storage servers while computations are performed on separate computational servers.
Therefore, data traverses through the network from the storage servers to the computational servers.
This creates a potential for accelerating queries computation by performing at least some of the calculations on the switches on-route to the computational servers.

In this work, we focus on sorting since it constitutes a fundamental part of queries answering.
Many queries require sorting, which is time and CPU consuming due to large data-sets and the complexity of sorting.
We aim to accelerate big data queries that require sorting by using the smart switch’s resources and computational power. 
We focus on the merge sort method since it is very common in databases due to its superior empirical run-time compared to other sorting methods.
We propose and examine a novel approach to accelerating queries at the server side by executing partial sorting at the programmable switch.
In our approach, we only partially sort the data, due to the limitations of programmable switches, which is enough to greatly reduce the run-time of completing the sorting task at the server.

\section*{Our contribution}
In this work we develop a partial merge sort algorithm that fits the popular PISA model adopted by most programmable switches, utilizing its multiple pipeline capability.
Specifically, we divide the data input into non-overlapping ranges.
Each pipeline in the switch handles its own range, and each value is passed to the server with its pipeline identifier.
This already creates a partial sorting, as the server can sort each range separately and then just concatenate the ranges by their original order. 

Further, a known observation about merge sort is that its running time depends significantly on how sorted the initial input is.
Specifically, merge sort is based on the notion of \emph{runs}, which are maximal sorted sub-sequences in the input.
The longer these pre-sorted runs are, the lower is their number, resulting in a faster sorting process.
For generating longer initial runs, we developed the MergeMarathon algorithm.
In MergeMarathon, instead of passing the data in its original order, we use the switch's programmable capabilities to change the stream's order, and output the data with longer runs.
This output is becomes the new input for the computing server; it is the same stream, but in a better sorted order.

MergeMarathon adds a few simple actions to each pipeline stage with a minor latency, following the switch's restrictions and limitations.
We evaluated the performance improvements that can be obtained at the computational server when utilizing MergeMarathon.
In our evaluation, we simulated various switch configurations and examined multiple datasets.
The results indicate that the use of MergeMarathon can reduce the run time of merge sort at the server side by 20\%--75\%.
We also represent trade-offs between the parallelism of the pipelines number and their length vs. latency and sorting run-time at the server.

\section{Background and Related work}
\label{sec:background}
\begin{figure}[t]
\begin{center}
\includegraphics[width=0.5\textwidth]{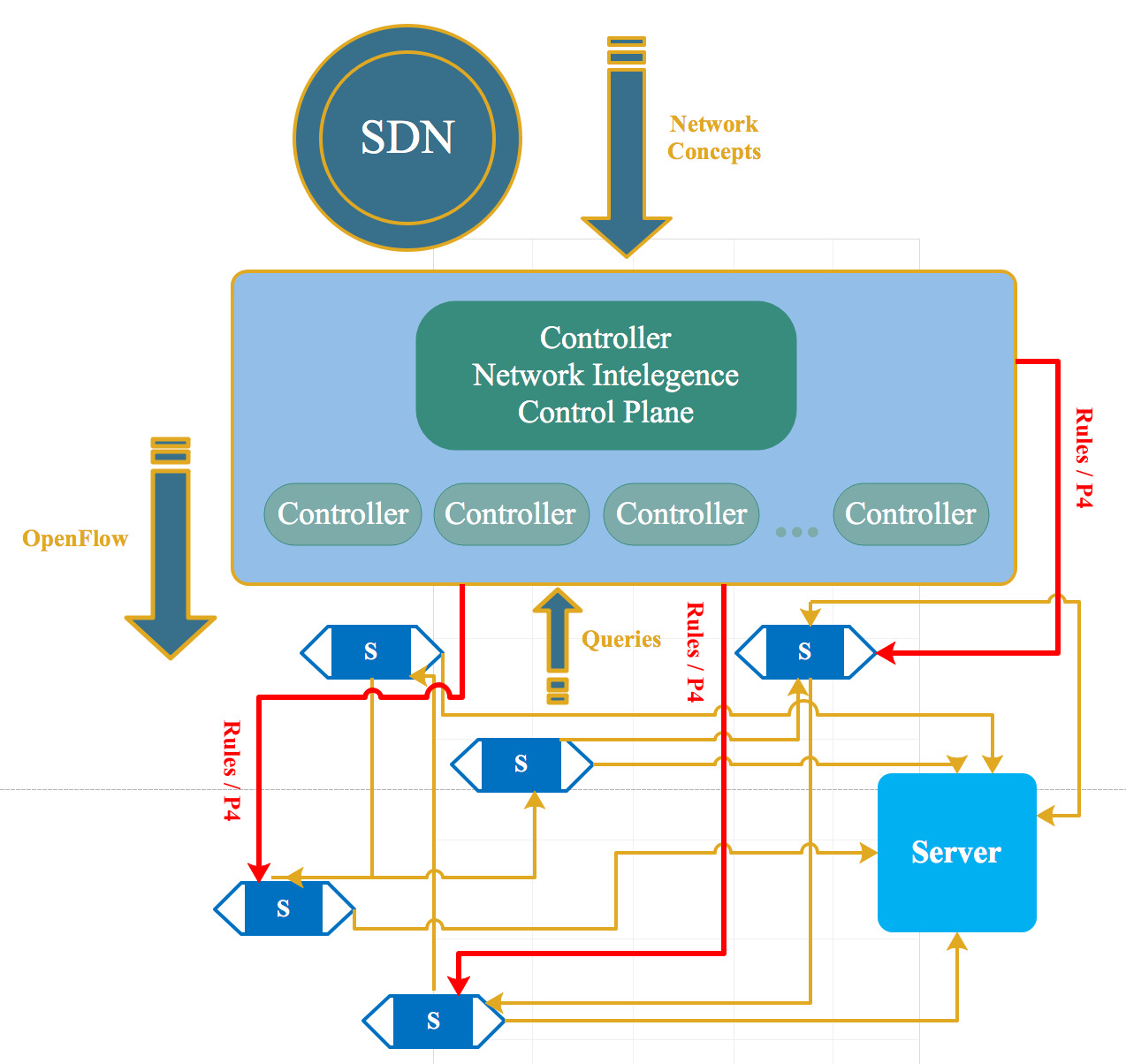}
\end{center}
\caption[Software Defined Network --- SDN]{Software Defined Network --- SDN with a controller separated from the network devices. Each device is configured with the rules that it received from the controller.}
\label{fig:s}
\end{figure}

In Software Defined Networking (SDN), the control plane is physically separated from the data plane~\cite{bosshart2013forwarding}.
Logic at the control plane dictates the forwarding policy and forwarding tables to the data plane (e.g., switches and routers) using an open interface, such as OpenFlow~\cite{mckeown2008openflow}.
This enables network controllers to determine the path of network packets across a network of switches. 
This network management approach enables to dynamically reconfigure the network in order to improve its performance, by using a controller that is separated from the switch hardware and enrolled as the intelligence of the network. 
Figure~\ref{fig:s} shows an illustration of the SDN architecture.

Programmable switches are at the state of the art of SDN, and have gained popularity in the last few years.
In particular, there has been significant development in programmable switches covering abstractions and platforms, dedicated programming languages and hardware as well as chips for such smart switches. 
In addition to routing packets, a programmable switch can also be programmed to perform various actions and applications such as collecting data from packets, analyzing it, compute certain statistics, etc. 

There is a natural trade-off between the speed of hardware and flexibility offered by programmablity of software that we need to pay attention to. 
For example, enabling the switch to perform a division action, which is quite complicated, would sacrifice the switch's speed and increase its latency.
To use programmable switches without sacrificing the speed significantly, the \emph{Reconfigurable Match Table} (RMT) abstraction was suggested~\cite{bosshart2013forwarding}. 
Together with the emergence of the \emph{Programming Protocol-Independent Packet Processors} (P4) language~\cite{bosshart2014p4}, the ability to combine the power of both software and hardware, and thus enjoy the benefits of each, has become feasible.

\subsection{RMT and PISA}
\begin{figure}[t]
\begin{center}
\includegraphics[width=0.8\textwidth]{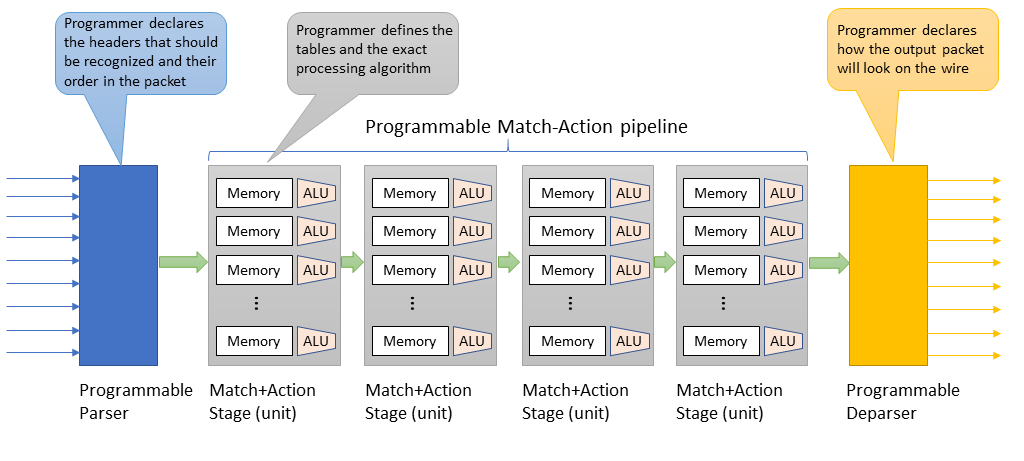}
\end{center}
\caption[RMT and \emph{PISA} model]{The path of an incoming packet in the programmable switch based on the RMT abstraction and PISA --- from the parsing stage through the match-action pipeline and exiting through the deparsing stage.}
\label{fig:pisa}
\end{figure}

RMT is a high performance abstraction for programmable switches. 
In this model, a packet goes through a parser that extracts its fields, then goes through an execution pipeline.
At last, is goes through a deparser, which packs the header and then the packet can go out to its next destination. 
Figure~\ref{fig:pisa} illustrates the path of an incoming packet in a programmable switch. 
Each stage in the pipeline is a match-action table, which first matches the packet, and can then modify it. 

\begin{figure}[t]
\begin{center}
\includegraphics[width=0.55\textwidth]{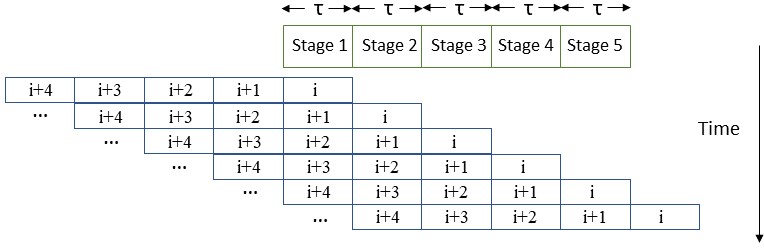}
\end{center}
\caption[Pipeline flow]{A timeline of the pipeline stages and processing. Every time unit, the packet leaves the current pipeline stage and moves to the next one, and another packet takes its place in the previous pipeline stage.}
\label{fig:pipeline}
\end{figure}

\begin{figure}[t]
\begin{center}
\includegraphics[width=0.95\textwidth]{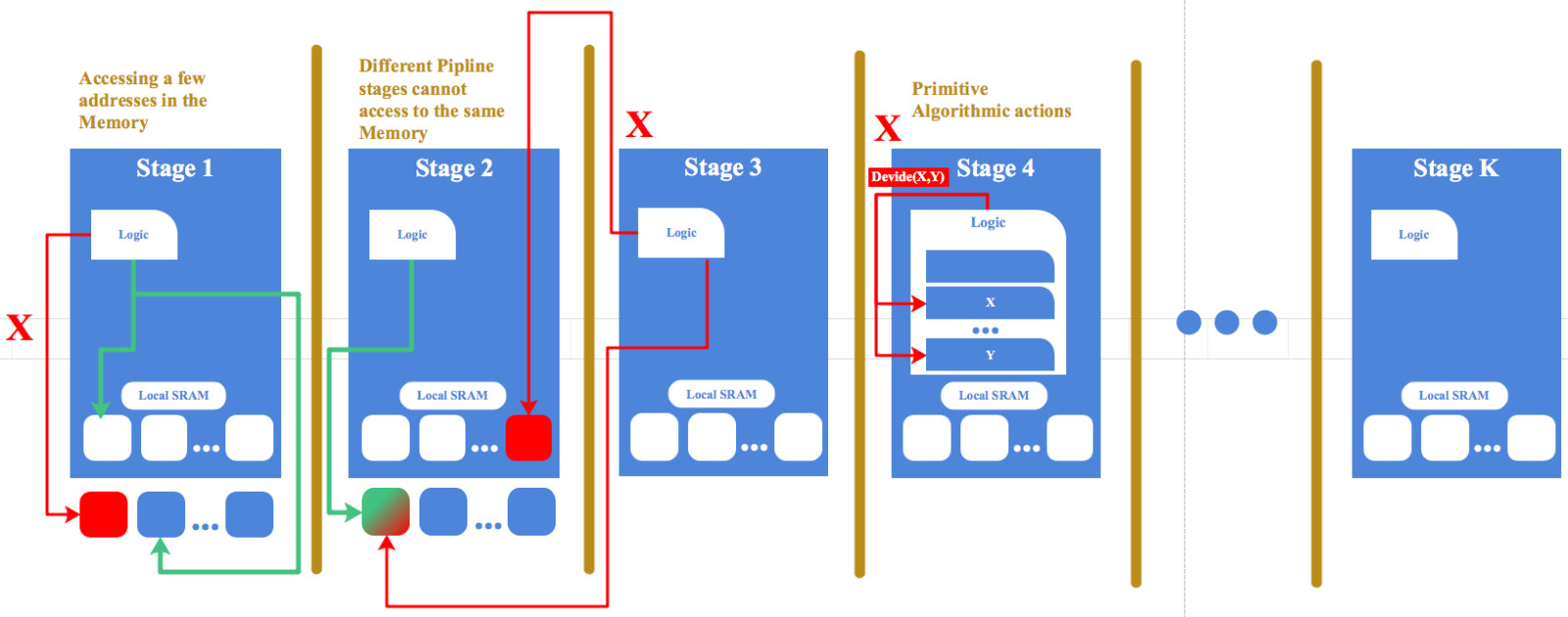}
\end{center}
\caption[RMT restrictions and limitations]{Each stage illustrates a different restriction or limitation of RMT: Limit the number of stages (k). Stage 1 shows that each stage has its own limited memory and addresses and cannot access the entire memory. Stages 2,3 are two different stages that cannot access the same memory region. In stage 4 the actions and branching are limited and complicated actions are not permitted.}
\label{fig:restriction}
\end{figure}

This model enables adding functionality to the switch, exploiting its pipeline and parallelism, without derogating the speed of packet processing, due to the following limitations and restrictions of this model:
First, there is a limited number of pipeline stages, to avoid long latency.
Second, each pipeline stage only has a small amount of static random access memory (SRAM) at its disposal to perform stateful processing. It also can access a few addresses in the memory region but not the entire memory as a result of the per-stage timing requirement.
Besides, to avoid the hazards of a pipeline architecture (packets are processed in parallel), two different pipeline stages cannot access the same memory region. There is only one particular pipeline stage that is allowed access to stateful memory blocks.
Third, each stage can perform only primitive arithmetic actions. 
Complicated actions, like division, are not supported. Also functions like finding a minimum or maximum in an array with many elements that spread on a wide part of the memory, are not supported. 
In general, the model does not support global compute functions.
In addition, branching is very limited within pipeline stages. 
Figure~\ref{fig:pipeline} illustrates a pipeline flow, to demonstrate the packets' processing that is executed in parallel.
Figure~\ref{fig:restriction} shows the restrictions on this process due to the RMT model.

\emph{Protocol Independent Switch Architecture} (PISA) is a an architecture for high-speed programmable packet forwarding that generalizes RMT. 
Programmable switches that follow the PISA model consist of multiple pipelines through which a network packet passes sequentially. 
These pipelines contain stages with disjoint memory that can perform a limited set of operations as the packet passes through them~\cite{dataplane,P4dataplane,tirmazi2020cheetah}. 
In PISA, a packet is parsed into individual headers. 
Headers and intermediate results can be used for matching and actions and can be modified, added or removed. 
At the end, the packet is deparsed, as depicted in Figure~\ref{fig:pisa}.

\subsection{PISA and P4}
While RMT is an abstraction for programmable switches, P4 is a novel high-level programming environment designed to be re-configurable, protocol independent, and target independent, and therefore it can use PISA as a target. 
P4 raises the level of abstraction for programming a network, offering an API that is oblivious to the actual hardware implementation of the switch~\cite{bosshart2014p4}. 
Because of its properties, P4 is very useful in network programming. 
Nevertheless, it also has some limitations and restrictions associated with programmable~switches.

Because P4 is not a Turing-complete programming language, and because of the restrictions and limitations of RMT for programmable switches, certain algorithms cannot be implemented at all with P4, while others may need some changes and adjustments to be implemented. 
For example, HashPipe~\cite{sivaraman2017heavy} is an algorithm that enables identifying the heavy hitter flows or flows with large traffic volumes in the data plane. 
It adapts Space-Saving~\cite{SpaceSaving} to P4.

Space-Saving is a counter-based approach that solves the top-k and frequency estimation problem. 
Counter based algorithms maintain a unique counter per flow. 
Space-Saving can return an estimation regarding which are the flow identifiers of the top $k$ flows in terms of flow frequency, and the flows that generate more than a given percentage of the overall packets. 
This, in addition to a per flow counting.
Space-Saving does so by maintaining a table with a limited number of counters, which are allocated to the flows with the highest count. 
Whenever a packet from a flow that has no counter arrives, the flow is given the minimal counter. 
When queried for the frequency of a flow, Space-Saving returns the value of the corresponding counter if it exists in the table, or the value of the minimum counter in the table. 
Space-Saving offers guaranteed performance, and was shown to be the best among similar approaches~\cite{SpaceSavingIsTheBest}. 

HashPipe enables identifying the heavy hitter flows or flows with large traffic volumes in the data plane. 
It implements a pipeline of hash tables in P4 which retains counters for heavy flows while evicting lighter flows over time. 
Packets may need to make two passes through this pipeline: once to determine the counter with the minimum value among $d$ slots, and a second time to update that counter. 
The second pass is possible through ``recirculation''~\cite{sivaraman2017heavy}.

Another example of P4 adaption is PRECISION~\cite{ben2018efficient}, an algorithm for top-k and frequency estimation, implemented with P4 and based on the RAP~\cite{basat2017randomized} algorithm, which is an improvement of Space Saving. 
RAP is a novel algorithm for the frequency and top-k estimation problems that is also based on counters.
RAP utilizes a probabilistic admission filter.
That is, a new flow receives a counter with some probability inversely proportional to the minimal counter. 
Thus, if the flow is heavy, eventually it will receive a counter, the one corresponding to the minimal monitored flow at that time. 
Light flows are not likely to receive a counter. 
In the original RAP, we need to find the minimum counter among all the monitored flows for calculating the probability of an un-monitored flow to receive a counter. 
The architecture of programmable switches does not permit finding (and replacing) the minimum element among all counters due to same-stage memory access restrictions~\cite{ben2018efficient}. 
For this reason, PRECISION also uses recirculation, i.e., if a packet is unmatched to any flow counter, it is probabilistically recirculated to claim an entry with the new packet’s flow ID. Notice that that not all probability formulas can be computed in P4, and hence they are being approximated.

Sequential Zeroing is a novel efficient algorithm in programmable switches with P4 for detecting heavy hitters on intervals. It is based on Modulu Sketch, an approach for intervals that makes each packet goes through several stages to update its counters. It works like a clock, as it zeroes the counter of the first stage when incrementing the counter of the second stage and so on~\cite{TOKKO20}. The Sequential Window algorithm with Modulo sketch uses several sketches to count the packets, and resets the oldest flow counter each time the whole sketch is full, to resemble a sliding view. Sequential Zeroing combines a zeroing approach of scheduling with the Sequential window. It works with sub-intervals, and removes outdated flow counts from the last sub-interval of the Sequential Window by scheduling. This is done by splitting the stages into sub-stages
and applying the scheduling.

\subsection{Databases and Sorting} 
Databases are at the core of many applications such as data warehousing and analytics~\cite{thusoo2010data}. 
In modern data centers, storage and computation are typically separated to different servers, and data needs to traverse the network from the storage server to the computation server.
With the growth of workloads, database systems are challenged to supply high performance for queries on large distributed data sets. 
Thus, modern query processing engines such as Spark SQL~\cite{armbrust2015spark} optimize query completion time by partitioning tasks to workers such that each worker processes only one data partition.
The results are then aggregated at a master~worker. 

Cheetah is a novel query processing system that partially offloads queries to programmable switches~\cite{tirmazi2020cheetah}.
Yet, it offloads only a part of the query rather than the enitre query. 
This way, it accelerates Spark SQL~\cite{armbrust2015spark} by pruning -- an abstraction to filter the data that comes from the workers. 
Then the master runs queries on the pruned dataset and the accepted results are the same as they would be on the complete dataset, but at much reduced completion time. 
For example, the switch can accelerate the DISTINCT query by removing some duplicates, and let the master remove the rest instead of all of them. 
Based on the pruning technique, additional algorithms have been developed for other more complex queries such as JOIN and GROUP BY.

The processing of many queries requires sorting, either because the query explicitly requests a sorted answer, or in order to greatly expedite the execution of complex operators.
For example, it is well known that JOIN can implemented in linear time if both relations are sorted~\cite{revesz2010introduction}.
Sorting is a very well studied problem, and many solutions have been proposed including, e.g., Quick Sort, Bubble Sort, Bucket Sort and more~\cite{cormen2009introduction}. 
In the area of databases, Merge Sort is a very popular algorithm, and is thus at the focus of our~research.

\section{Problem Statement and Approach}
\label{sec:problem}

As mentioned previously, many database queries require sorting, and sorting is a CPU intensive task. 
Our goal is to accelerate sorting by pre-processing the input for the server using the power of programmable switches.
We focus on Merge sort due to its popularity in databases.
For completeness, we start by restating the Merge sort algorithm and then provide intuition on how programmable switches might help.

\subsection{Merge Sort}
The main idea behind Merge sort is to inductively merge short sorted sub-sequences, called \emph{Runs}, into larger ones, until the entire sequence is sorted.
Since Runs are sorted, their merging process takes linear time.
The fact that Runs are merged in the order they are encountered translates into a more virtual memory friendly access pattern compared to, e.g., Quick sort, which is why Merge sort is so popular in databases.
More~formally,
\begin{definition}
A \emph{Run} is a maximal ascending sequence.
\end{definition}

\begin{algorithm}
\caption{Merge sort of order $k$:}
\label{alg:merge}
\begin{algorithmic}[0]
\State Repeat
\State \quad Stage 1:  Divide the input into sets of Runs, such that each set contains $k$ Runs 
\State \quad\quad\quad\quad\quad\quad\quad\quad\quad\quad\quad\quad\quad\quad\quad\quad\quad\quad\Comment{the last one may contain fewer than $k$}
\State \quad Stage 2: Merge each set of Runs into a single Run
\State Until there is one Run left
\end{algorithmic}
\end{algorithm}

\begin{figure}[t]
\begin{center}
\includegraphics[width=0.7\textwidth]{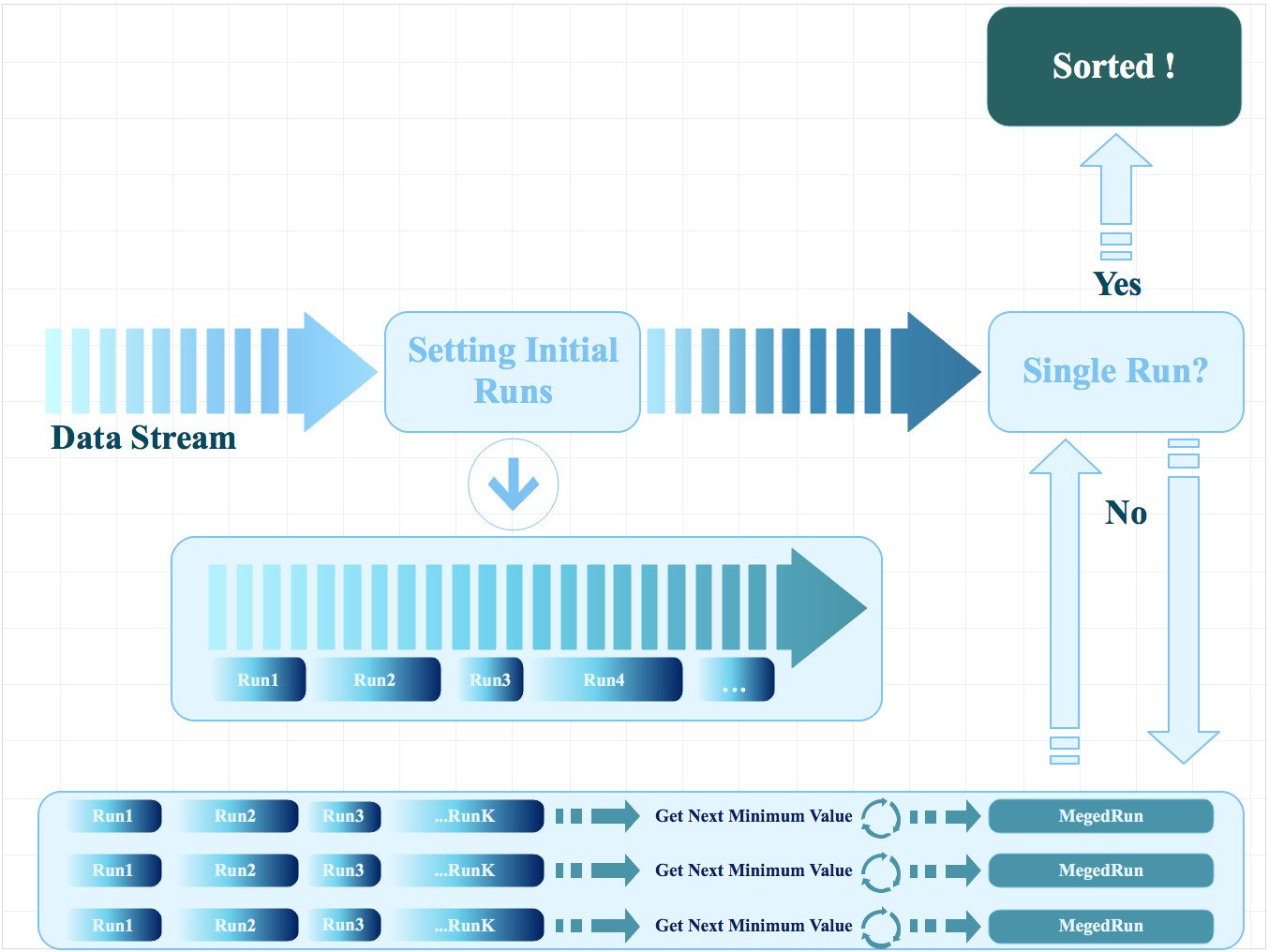}
\end{center}
\caption[Merge sort flow]{Merge sort flow: Data stream is divided into sets of $k$ initial Runs. Each $k$ are merged into one single Run. Then the remained merged Runs are divided into sets of $k$ Runs and merged. This process is repeated until there is only one Run that constitutes the whole data stream sorted.}
\label{fig:merge_sort}
\end{figure}

\begin{figure}[t]
\begin{center}
\includegraphics[width=0.7\textwidth]{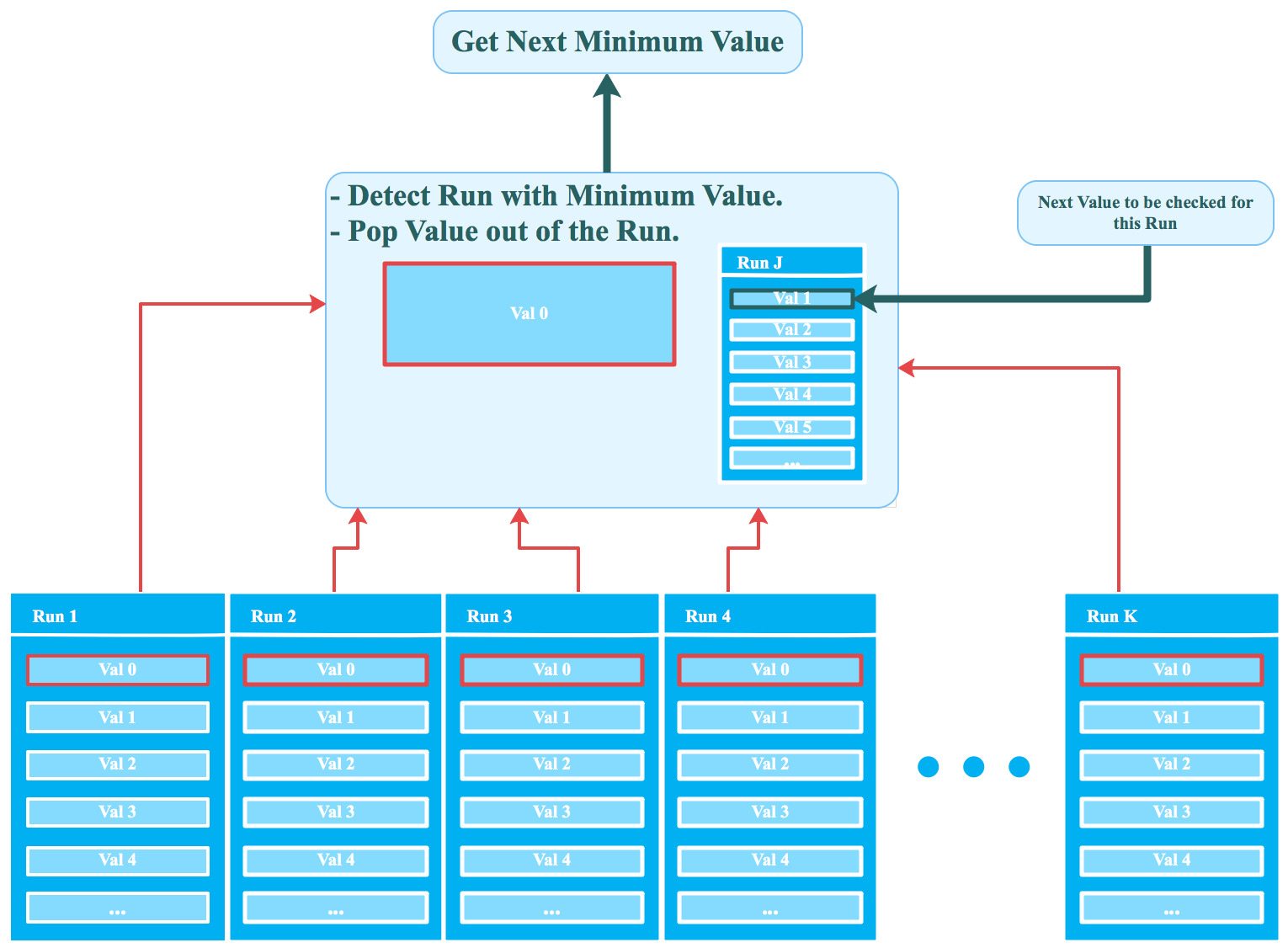}
\end{center}
\caption[Merge iteration]{A closer look on merging the $k$ Runs. By definition, each Run is sorted. Therefore, the first element of the new merged Run is the minimum among the first element of each Run. After detecting the minimum, it is removed from the elected Run and the second value in this Run becomes the first. To determine the next element in the merged Run, the minimum among the first element of each Run is detected again. This process is repeated until there is only one single merged Run. This process's run time is linear in the combined length of the merged Runs.}
\label{fig:merge_sort_inside}
\end{figure}

Stage 2 of Algorithm~\ref{alg:merge} can be done in parallel over all sets of Runs. 
Notice that the the overall run-time of Merge sort depends on the number of Runs logarithmically. This is because each stage reduces the number of Runs by $k$. 
Thus, if we look at a given input, longer Runs translate into fewer Runs, and therefore the run-time of merge sort becomes faster the longer the initial Runs are.
Figure~\ref{fig:merge_sort} illustrates the process of merge sort in general while Figure~\ref{fig:merge_sort_inside} illustrates the internal merge process in more details.

\subsection{Expediting Merge Sort by Utilizing a Programmable Switch}
In a typical distributed database deployment, the main function of the switch is simply to forward packets to a computational server, and the server is the one executing the sort.
Under the assumption that the packets traverse the switch in their original order, the time and resources it takes for the server to execute merge sort depends on the original input.

\subsubsection{Main contribution}
In our work, we expedite the sorting process at the computational server by assigning some pre-processing to the switch, which occurs while the packets traverse through the switch on their way to the server. 
Specifically, we use the pipeline and match-action table, with the rules of the RMT abstraction mentioned before, to manipulate the order of the input packets and output them partially ordered to the server. 

To that end, we employ two ideas:
First, we divide the input into ranges, where each pipeline segment in the switch takes care of a different range of the input domain.
The ranges are non-overlapping to each other and together cover the entire domain.
This way the server only needs to sort the output of each of the segments and then concatenate the sorted sub-streams into a completely sorted relation.
This expedites the execution both because each sort at the server runs on a shorter range, and because each sub-stream fits better into virtual memory and the hardware cache.

The overall time complexity of Merge sort is $O(N*log(N))$. 
If a stream with $N$ values is partitioned into $S$ equal size ranges, the number of sub-streams is $\frac{N}{S}$, and thus, the overall time complexity of Merge sort is $O(S*\frac{N}{S}*log(\frac{N}{S})) = O(N*log(\frac{N}{S}))$, assuming the sub-streams are sorted sequentially and not in parallel.

\begin{figure}[t]
\begin{center}
\includegraphics[width=0.7\textwidth]{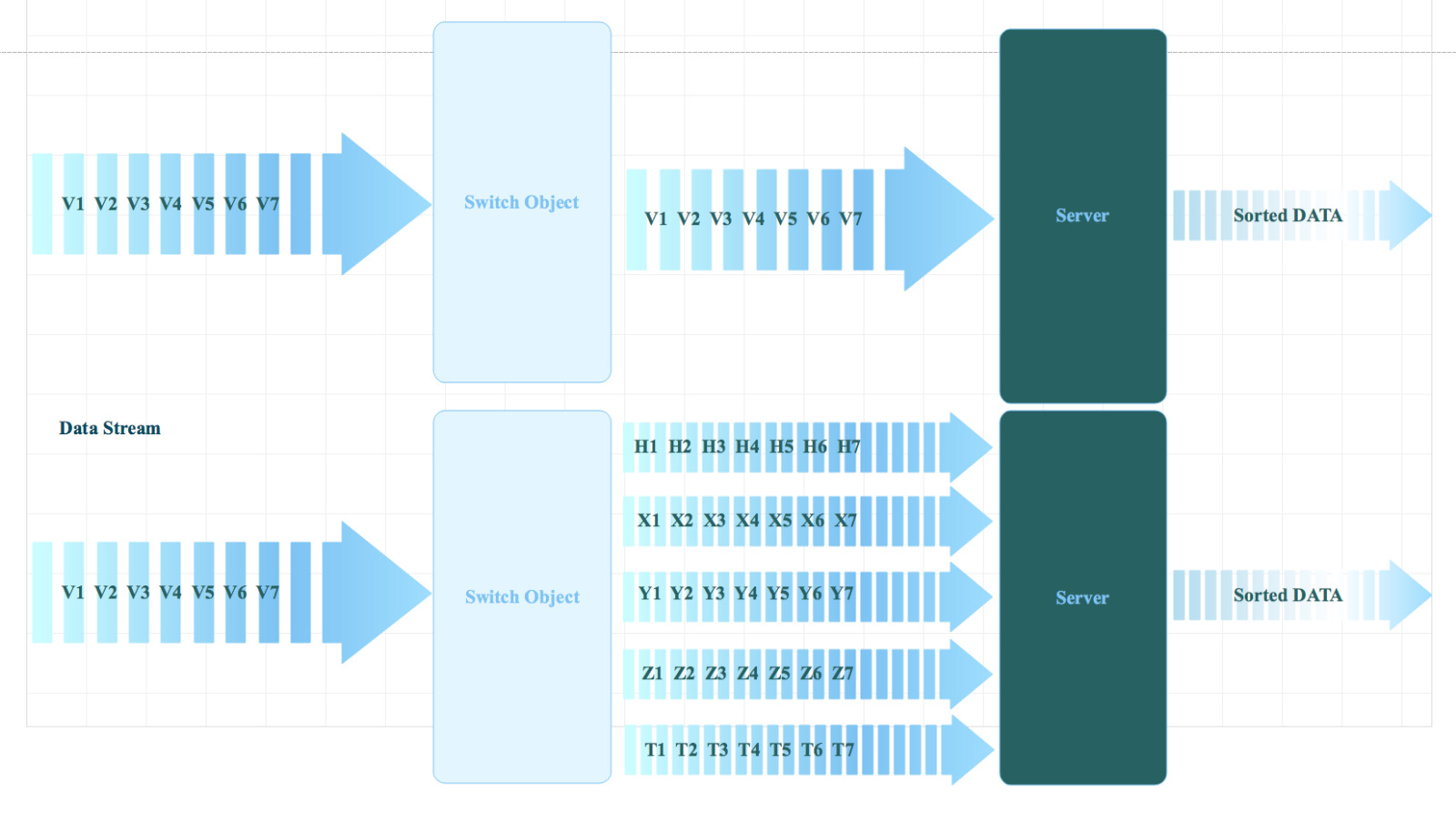}
\end{center}
\caption[Sorting with and without our algorithm]{In the upper illustration, the input passes through the switch and arrives at the server in its original order. In the second illustration, the input is divided into ranges and undergoes a partial sorting, before arriving at the server.}
\label{fig:withorwithout}
\end{figure}

Second, we use the pipeline segment to store the packets temporarily in an ascending order and thus we create longer initial Runs in the input (compared to their original order). 
Obviously, the longer the initial Runs are in the input, the faster Merge sort would execute in the server as it would require fewer sorting and merging stages, as illustrated in Figure~\ref{fig:withorwithout}.

To better understand the execution times, we state a formula based on the number of initial Runs and their average length. 
Here we assume that stage 2 of the Merge sort algorithm is done in parallel. 
We denote:
\begin{description}
\item[$N$]--- Number of elements in the input.
\item [$\tilde{r}_{\textnormal{init}}$]--- Average length of initial runs.
\item[$\ell=\frac{N}{\tilde{r}_{\textnormal{init}}}$]--- Number of initial Runs.
\item[$k$] --- Merge sort order.
\item [$\tilde{r}_{\textnormal{i}}$]--- Average length of Runs in the $i^{th}$ iteration.
\end{description}
Consider the first partition to Runs: 
$
\underbrace{
\underbracket{\tilde{r}_{\textnormal{init}}}_{1},
\underbracket{\tilde{r}_{\textnormal{init}}}_{2},
\underbracket{\tilde{r}_{\textnormal{init}}}_{3},
\ldots,
\underbracket{\tilde{r}_{\textnormal{init}}}_{\ell}}_{\ell}
$

From the structure of the Merge sort algorithm, it is clear that the number of iterations needed to sort the input is $\log_k\ell$.
Under the assumption that each $k$ runs are merged in parallel, the time each iteration of merging $k$ runs takes is determined by the average length of the Runs multiplied by the number of Runs. 
We denote it by $k\cdot\tilde{r}$. 
We also notice that the average length of runs in the $i^{th}$ iteration is $\tilde{r}_{i}=k\cdot r_{i-1}$, when $i>0$. 
We now compute the merge time:
\begin{description}
\item[iteration 1]: $k\cdot\tilde{r}_{\textnormal{init}}$
\item[iteration 2]:
$k\cdot(k\cdot\tilde{r}_{\textnormal{init}})=k^2r_{\textnormal{init}}$
\item[$\dots$]
\item[iteration $i$]:
$k\cdot(k\cdot\tilde{r}_{\textnormal{i-1}})=k\cdot(k\cdot(k\cdot\tilde{r}_{\textnormal{i-2}}))=...=k^i\cdot{r_{\textnormal{init}}}$
\end{description}
By summing the time of all iterations, we get the following formula for merge sort:
\[\sum_{i=0}^{\log_k\ell}k^ir_{\textnormal{init}}\]

Notice that the smaller $i$ is, i.e., fewer iterations required to completely sort the input, the faster the run-time of Merge sort becomes.
To reduce $i$, we can reduce $\ell$ by increasing $\tilde{r}_{\textnormal{init}}$.
Our work proposes a method of increasing $\tilde{r}_{\textnormal{init}}$ inside the programmable switch.
Notice that even if in each iteration the merge of each $k$ Runs is done sequentially and not in parallel, the number of the iterations remains unchanged. 
The time of each iteration becomes fixed and equals to $N$.

\subsubsection{The Design Trade-off}
In our work, we analyze the impact of both the number of segments in the switch, i.e., the level of parallelism, and the number of pipeline's stages of each segment. 
On one hand, more segments and stages mean more hardware and also the latency depends on the length of the pipeline.
On the other hand, an increased number of segments reduces $N$ while the number of stages linearly impacts $\tilde{r}_{\textnormal{init}}$.

\section{Implementation}
\label{sec:implement}
Our implementation includes a framework that simulates a programmable switch, obeying the restrictions and limitation of RMT and the PISA model. 
It also includes a server that executes the sorting. 
We implemented the infrastructure and algorithm in C language, following the rules of RMT.
Thus, it can be converted to P4 for a real programmable switch.
One can configure our switch simulator to fit various programmable switches capabilities, with different hardware, infrastructure and configurations to explore their performance~impact.

\subsection{Programmable Switch Part}
Our application simulates a programmable switch and follows the PISA model and RMT rules. 
The switch part contains multiple Match-Action tables, each with multiple entries.
This is simulated by a variable number of pipelines, each with a variable number of stages.
We refer to each pipeline row as a segment.
The switch is represented by a data structure that is configured by the user.
The user passes the number of segments and their length as a parameter, which determines the switch's structure, thus can simulate a real and available programmable switch.

The switch can be configured to work with a domain of values.
In our implementation, the switch supports integers, but can be easily adapted to other domains.
The maximal value the switch can support is an optional argument for the user whose default is the maximal value of integer type. 

The switch also maintains the segments structure. 
Each segment is represented by an array, where each cell of the array represents a pipeline stage.

\subsection{Server}
This application simulates a server that receives data from the switch and applies the merge sort algorithm, as described in Section~\ref{sec:problem}.
The server has ports, one for each switch' segment, and we configure it to accept a different segment from each port, thus it can merge each port separately and reduce the computing time.

\subsection{How it All Works}
 
\subsubsection{Switch}
\begin{figure}[t]
\begin{center}
\includegraphics[width=0.9\textwidth]{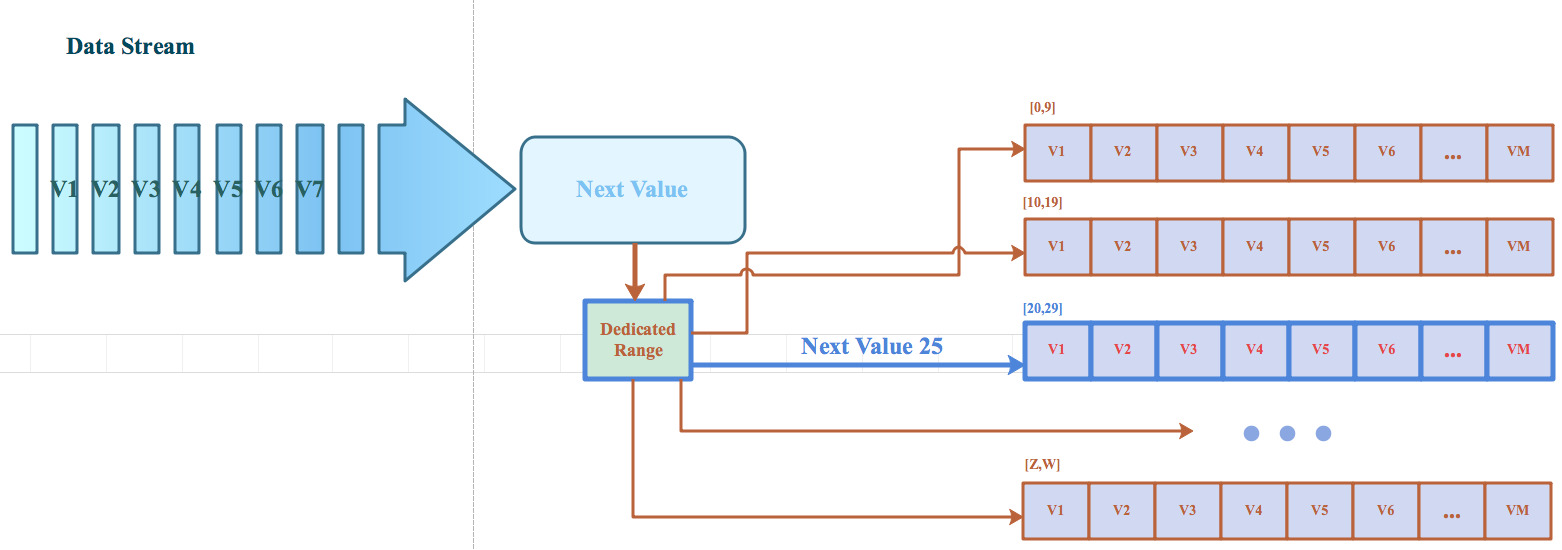}
\end{center}
\caption[Finding the right segment]{Finding the right segment --- Each value that arrives is directed to the pipeline determined by its range. The right range is found by a simple comparison actions.}
\label{fig:findsegment}
\end{figure}

The switch is initialized with the arguments provided by the user: number of segments, segments' length and the maximum value.
The switch divides the segments to ranges and all segments are initially empty (i.e., contain initial values that are outsize the domain's boundaries).
The switch receives the packets' stream as a text file, and starts to read line after line. 
We assume that the packets' header are already parsed (we can assume that because every switch has a parser that parses the packets' header) and thus each line represents a packet from the stream. 
When the packets arrive one after another, each value (row) is directed to the suitable segment according to its range. 
Because the ranges are non-overlapping and together cover the entire domain, it is guaranteed that there is exactly one such range.
See illustration in Figure~\ref{fig:findsegment}.

After finding the right segment, the packet starts propagating in the pipeline stages to find its correct stage to be stored.
The pipeline stages store values in ascending order, and the new value is located in the right place, while all the bigger values are bubbled one stage forward.
When the stage is full with values, the switch starts to emit old values from the pipeline, by preserving the order in the pipeline.
Each time a new value arrives, an old value is evacuated, and the new one is inserted to the fit place.

\begin{figure}[t]
\begin{center}
\includegraphics[width=0.6\textwidth]{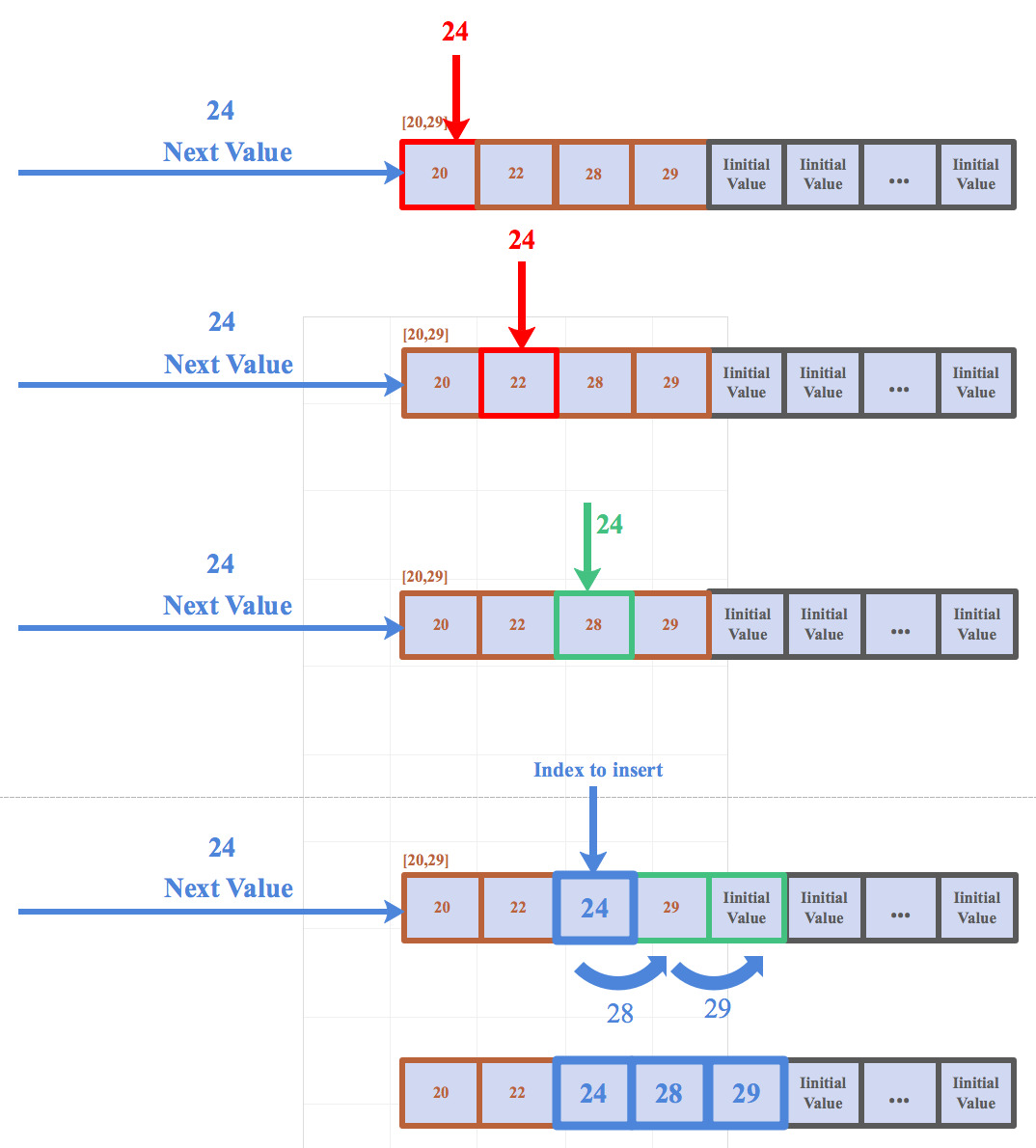}
\end{center}
\caption[Insert to a non-full segment]{A value insertion in case the segment is not full yet. The value should be inserted to the third index, thus they are swapped. After the swapping, all the values after the swapping index move one stage forward.}
\label{fig:insertinit}
\end{figure}

\begin{figure}[t]
\begin{center}
\includegraphics[width=0.6\textwidth]{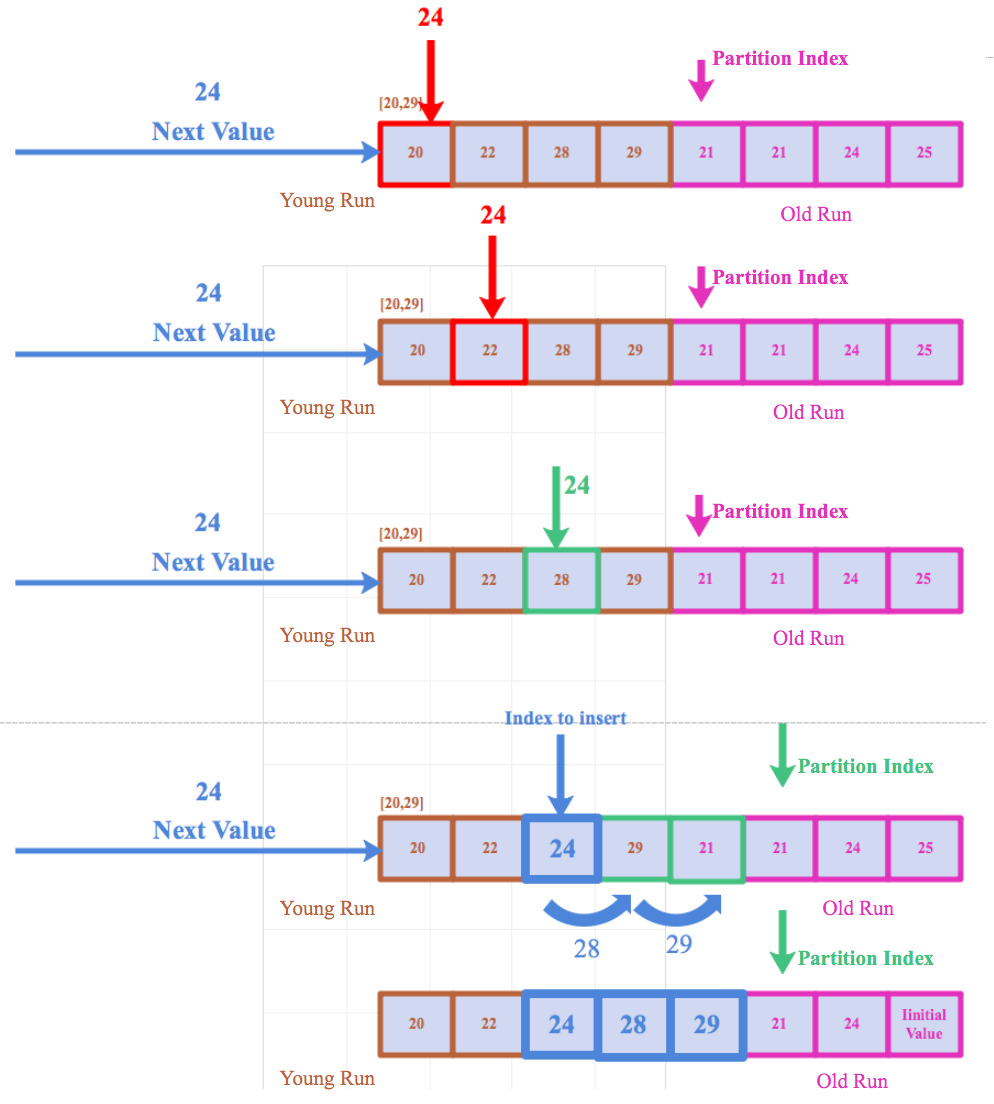}
\end{center}
\caption[Insert to full segment]{A value insertion in case that the segment is full. The first value of the older Run is removed, and the new value finds its place among the younger Run values. The values in the stages following the chosen stage move one stage forward while the last value in the younger Run takes the place of the value that has been removed. The start of the older Run is also advanced one stage forward.}
\label{fig:insertful}
\end{figure}

This way, the segment holds either a single Run or parts of two Runs at any given point.
In the latter case, the older Run shrinks while the younger one grows respectively. 
When the last value in the older Run is removed, the older Run no longer occupies any space in the switch.
At this point, the younger Run becomes the older Run and the next incoming value creates a new younger Run. 
In more details, we distinguish between the following~cases:
\begin{description}
\item[Case 1:] The segment is empty, meaning that the value is stored in the first stage.
\item[Case 2:] The segment is not empty, but not full either. Here, the new value is compared to the values that are already stored in the pipeline (when the pipeline is not full, the first stages are populated). For each stage, if the current value is bigger or equal to the value in the current stage, we continue to the next stage. If the current value is smaller than the current stage, then the current value and current stage are swapped.
In this case, the new current value is passed to the next stage, and they are also swapped.
This process is repeated for each stage, until there is an empty stage, and the last current value is stored in that stage.
If the current value is bigger or equal to the value in the last populated stage, it is stored in the first available stage.
See illustration in Figure~\ref{fig:insertinit}.

\item[Case 3:] The segment is full.
The first time this happens, all the values inside the segment are ordered in ascending order and constitute a Run.
The smallest value is moved to the output, and the new current value substitutes it, creating a new Run.
Now there are two Runs in this segment.
In fact, the latter happens each time there is a single Run in the segment (which happens every segment length arrivals).

If there are two Runs, the smallest value in the older Run is removed. 
That is, the first index of the older Run becomes available and assigned to be the last available index of the younger Run. 
Now the new current value is inserted to the younger Run in the same way as in case 1 with respect to the last new index of the younger Run. 
The new first index of the old Run is forwarded to the next stage. 
When the old Run is empty, the younger Run becomes the older Run and the first index of the older Run is the first index of the previous younger Run, which is also the first index of the pipeline.
See illustration in Figure~\ref{fig:insertful}.
\end{description}
 
Each value is emitted with its matching segment number.
When there are no more lines to read, the switch does a flush action and outputs the remaining values. 
It first does one pass and flushes out the older Run.
Then, it performs a second pass and flushes out the younger Run (if exists).

\subsubsection{Server}
The server receives the values and their segments. 
It invokes the merge sort algorithm on each segment separately (every two values in two different segments are strangers). 
After each segment is sorted, the server unites the segments according to their serial number, and then outputs all the values into one output. 
The final output is the sorted values in ascending order.

\section{MergeMarathon Algorithm}
The pseudocode of our algorithm, MergeMarathon, is listed and explained below.
First, we state two data structures, one for the switch and one for the pipeline segments.
\begin{lstlisting}
structure Switch{
  int number_of_segment;
  int segment_length;
  int max_value;
  Segment* segment;
}
\end{lstlisting}

\begin{lstlisting}
structure Segment{
  tuple range;
  int last (last populated index);
  int partirion_index;
  stages[segment_length];
}
\end{lstlisting}

\begin{algorithm}
	\caption{MergeMarathon}
    \begin{algorithmic}[1]
		\State input: number\_of\_segments, segments\_length, max\_value, stream of integers {$>0$}.
		\State output: Stream of runs.    	
		\Statex
	    \Procedure {$Set Ranges$}{$Segment$}:
            \State Set $q$ to be $max\_value / number\_of\_segments$.
            \State Set $r$ to be $max\_value \mod number\_of\_segments$.
            \For{$i=0$ to $r$}
                \State Set {$Segments[i].range = [(q + 1)*i, (q+1)*i + q]$}
            \EndFor
            \State Set {$T = r*(q+1)$}
            \For{$i=r$ to $number\_of\_segments$}
        	    \State Set {$Segments[i].range = [T + (i - r)*q, T + (i + 1 - r)*q]$}
        	\EndFor
        \EndProcedure
        \Statex
		\Procedure {$Switch Initialize$}{}:
		    \State Set $number\_of\_segment$.
		    \State Set $segment\_length$.
		    \State Set $max\_value$.
		    \ForEach{$s \in Segments$}
		        \State $Set Ranges$ {$(S)$}
        	\EndFor
    	\EndProcedure
    	\newcounter{lastlinenum}
    	\setcounter{lastlinenum}{\value{ALG@line}}
    \end{algorithmic} 
\end{algorithm} 

\begin{algorithm}[tp!]
	\caption{MergeMarathon - Continued}
    \begin{algorithmic}[1]    
    \makeatletter
    \setcounter{ALG@line}{\value{lastlinenum}}
    \makeatother
        \Procedure{Segment Insert Value}{$s$, $v$}:
            \If{there is available $stage \in s.stages$}:
                \State Insert $v$ to suitable stage:
                \State Set {$x$ = $s.stages[last].value$}
                \If{$v > x$}:
                    \State Set {$s.stages[last+1].value = v$}
                \Else
                    \State Find first $i$ such as: Set {$x = s.stages[i].value$}, {$x > v$}.
                                    \ForEach {$i < j \leq last$}:
                    \State Set{$s.stages[j].value = s.stages[j - 1].value$}
                \EndFor
                \State Set {$s.Stages[i].value = v$}
                \EndIf
            \Else
                \State {$Output Stream \leftarrow s.stages[partition\_index].value$}
                \If{$s.partition\_index = 0$}:
                    \State {$s.stages[partition\_index].value = V$}
                \Else
                    \State Set {$x = s.stages[partition\_index - 1].value$}
                    \If{$v \geq x$}:
                        \State Set {$s.stages[partition\_index].value = v$}
                    \Else
                        \State Find first $i$ such as: Set {$X = s.stages[i].value$}, {$x > v$}.
                        \For{$each$ $i \leq j \leq s.partition\_index$}:
                            \State Set {$segment.stages[j].value = s.stages[j-1].value$}
                        \EndFor
                        \State  Set {$s.stages[i].value = v$}
                    \EndIf        
                \EndIf
            \EndIf
            \State {$s.partitions\_index = (s.partitions\_index + 1) \mod segment\_length$}
        \EndProcedure
        \Statex
    	\Procedure{$Switch Insert$}{$v$}:
    	    \State Find {$s \in Segments$} such as: $Segment.range = [a, b]: {v \geq a} and {v \leq b}$.
            \State {Segment Insert(s, v)}.
        \EndProcedure
        \Statex
    	\Procedure{$Switch Flush$}{}:
    	    \ForEach{$s \in Segments$}
    	        \ForEach{$stage \in s.stages[stage].value$}
		            \State {$Output Stream \leftarrow s.stages[stage].value$}
        	    \EndFor
        	\EndFor
        \EndProcedure
    \Statex
    \Procedure{$Apply Switch$}{input}:
        \State Switch Initialize();
    	\ForEach {$v \in InputStream$}:
        	\State  {$Switch Insert$}($v$)
    	\EndFor
    \State Switch Flush();
    \EndProcedure
	\end{algorithmic} 
\end{algorithm} 


\subsection{MergeMarathon in the PISA Model}
We now explain how each phase in the algorithm can be implemented in the PISA model.

\subsubsection*{$Switch Initialize$}
The number of segments and their length are part of the architecture of the switch. 
In our work it is a parameter, in order to explore the impact of this design choice on the performance of the overall sorting task. 
The maximum value is needed only for the ranges calculation at initialization.

\subsubsection*{$Set Ranges$}
To set the ranges, the algorithm executes one division action and saves the remainder. 
The programmable switch cannot perform a division in the RMT model. 
This can be solved in several ways: 
The results can be calculated at the server before the beginning of the execution and then dictated to the switch. 
They can also be received as a parameter, or they can be calculated by approximation of division with TCAM and log operations~\cite{tirmazi2020cheetah}.
This is feasible because the division is done only once at the beginning.

\subsubsection*{$Switch Insert$}
Finding the right segment for an item can occur at the parsing stage, before the algorithm starts.

\subsubsection*{$Segment Insert$}
This part is the heart of the algorithm. 
When a packet arrives, it is parsed and traverses through the pipeline stages as part of the routing process.
The value that is parsed is compared each time with the value in the current stage. 
Comparison is a simple action that is allowed. 

In each stage, when the packet arrives, a comparison is performed.
If needed, the value in the packet swaps with the value in the stage. 
The bubbling of the values forward is performed by placing this value in the packet. 
In particular, to swap values, the value in the stage is swapped with the one in the packet, and then the packet moves on to the next stage. 

The output value is saved in the packet when we first identify it. 
It is output with the segment's number as a port number, and thus the server can identify the segment. 
We notice that each stage has its own memory and there is no branching or access to memory of any other pipeline stage. 
Also, this requires only a small amount of~memory.

Since the packets are processed in parallel, and each stage in the pipeline processes another packet, we cannot afford a partition index that is implemented by shared memory.
This is overcome by a dedicated variable that signals whether the stage is the partition index or not. 
This variable counts each packet that passes through the stage. 
From the algorithm's structure, after the first $segment\_length$ iterations, the $i^{th}$ stage becomes the partition index every time the iteration counter modulo $segment\_length$ equals to $i$ (when the counting of the stages and iterations starts from 0). 
It is enough that each stage detects if it is the partition index itself, because the comparison and swapping are performed locally in the stage.
When a packet traverses through the pipeline, it can be swapped with the stages until it passes the partition index stage.
When it arrives to the partition stage, it is marked, so the next stages can notice that this packet must not be swapped anymore.

In the first segment length iterations, there are still empty stages (with the initial value).
According to our algorithm, each stage can maintain a Boolean flag bit to mark stages with the initial value.
When a packet passes through an initial value stage, its value can be stored in that stage.

All the processing can be done in parallel, packet by packet.
Each stage can process its temporal packet because each stage knows its value and the current packet value, so the unit can perform a comparison.

\subsubsection*{$Switch Flush$}
After the input stream ends, the pipeline stages are still full with values. 
To emit these values while preserving their order, we perform a flush action. 
The Flush can be done by a particular packet that passes through the pipeline and collects the values in their order at the switch. 
The flush can be finished in two rounds using the recirculation mechanism of P4: 
In the first round, the packets of the old Run are flushed while items belonging to the young Run are ignored (to continue the sequence). 
In the second round, it flushes out the younger Run.

\subsection{Requirements and Assumptions}

As described in Chapter~\ref{sec:background}, the storage and computation functions are assumed to be hosted on different servers. We place a few assumptions and requirements to enable our algorithm to be realized in a real data center system. 

Below, we refer to the storage server, the computation server and the programmable switch.
First, in our algorithm each packet contains one record. That is, we assume that the application that sends the data to the computation server, sends it with one record per packet. However, in a case that each packet contains more than one record, if the number of such records is small, fixed, similar for all packets, and known, our algorithm can be adjusted to still work, e.g., by placing copies of the keys in the header.

Second, the computation server should inform the storage server what is the key for the sort.
This way, the storage server can build the packets accordingly.
In particular, the packet header should contain the sorting key, in order to supply the programmable switch the ability to look at the key and change the order of the packets.

Third, due to our algorithm, the switch outputs each value (packet) with its corresponding segment number (pipeline). The computation server should support the different segment concept and sort each segment separately. The switch can use a particular flag in the packet, and the server can read the flag and associate the value to the right sub-stream.

\section{Performance Analysis}
\label{sec:perf}
As mentioned before, we implemented our simulator in C. 
Our implementation includes a programmable switch part that performs our MergeMarathon algorithm, and a server that receives the output of the programmable switch as an input in the new order computed by the switch.

We performed multiple experiments to explore the algorithm's performance under three different representative traces: random trace, network trace and memory access trace.
The first is a randomly generated trace with uniform distribution that includes 100M numbers.
The second is a real network trace from CAIDA\footnote{\url{https://www.caida.org/data/passive/passive_dataset_download.xml}}. 
We parsed the trace and took the length of the packet in each line, preserving their original order. 
Thus, we created a trace of 100M numbers.
The third is a trace of memory accesses for I/O with sizes from SYSTOR~17 in the SNIA repository\footnote{\url{http://iotta.snia.org/traces/4931}}. 
We parsed this trace to extract only the sizes while preserving their original order, and created a trace of numbers with 77M lines. 

For each of the traces, we first ran it in the server only (without MergeMarathon) and measured the time for performing the sorting by Merge sort.
This experiment examines the existing situation in which a regular switch only passes the packets to the server, and the server performs the entire sorting process itself.
In the second set of experiments, we checked various switch architectures with various combinations of number of segment and segment lengths. 
In all the experiments, we set the order of Merge sort $k = 10$.
This way, we studied the influence of parallelism (multiple pipelines) compared to pipeline length.

We expect that an increase in the segments number will result in the run-time of merge sort being reduced.
We expect the same for the length of segments. 
We checked if there are optimal values for the number of segments and their length.
That is, a point that provides good performance, yet the improvement with larger values is insignificant.
We also collect statistics about the output of the switch in each variation, such as average number of Runs per segment, to better understand the influence of the length on the run-time.

\subsection{Without MergeMarathon}
\begin{figure}[t]
\begin{center}
\includegraphics[width=0.4\textwidth]{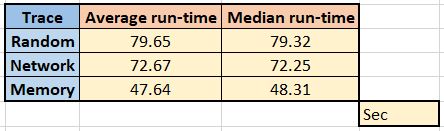}
\end{center}
\caption[Server alone results]{The results of average and median run-time in seconds for each one of the traces, without MergeMarathon. The random trace and the network trace contain 100M values while the memory trace contain 77M values.}
\label{fig:serveralone}
\end{figure}

First, we checked the run-time for Merge sort without MergeMarathon.
The server performed Merge sort on the original input so the initial Runs depend on the original order.
We performed 10 identical experiments for each trace, and measured the run-time.
We calculated the average and the median time among those experiments.
The results are listed in Figure~\ref{fig:serveralone}.

\subsection{With MergeMarathon}

In the next experiments, we set the switch in our simulations to $x$ pipelines with $y$ stages each, where $x\in \{{1, 4, 8, 16, 32, 64, 128}\}$ and $y\in \{{4, 8, 16, 32, 64, 128}\}$.
That is, for each switch configuration with $x$ pipelines, the length of all the pipelines in the switch is $y$ stages.
For each configuration, we ran all 3 traces.
Each run was executed 10 times. 
We calculated the average and the median time of those experiments.

\subsubsection{Exploring the Performance in 3D}
\begin{figure}[t]
     \centering
     \begin{subfigure}[b]{0.49\textwidth}
         \centering
         \includegraphics[width=0.9\textwidth]{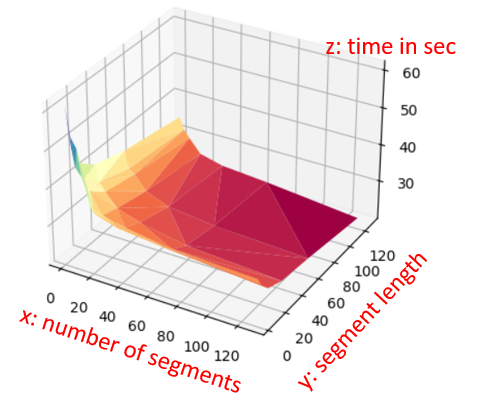}
         \caption{$average$}
         \label{fig:randomaverage1}
     \end{subfigure}
     \hfill
     \begin{subfigure}[b]{0.49\textwidth}
         \centering
         \includegraphics[width=0.9\textwidth]{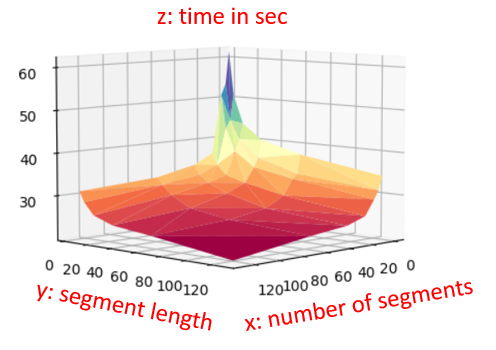}
         \caption{$average$}
         \label{fig:randomaverage2}
     \end{subfigure}
     \hfill
     \begin{subfigure}[b]{0.49\textwidth}
         \centering
         \includegraphics[width=0.9\textwidth]{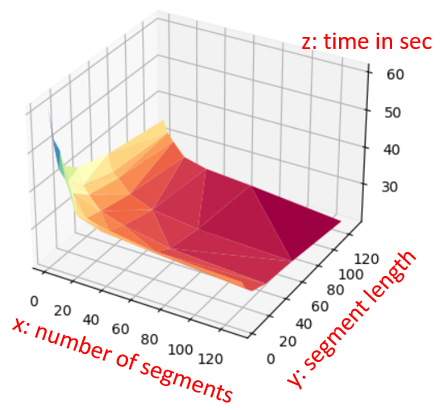}
         \caption{$median$}
         \label{fig:randommedian1}
     \end{subfigure}
     \hfill
     \begin{subfigure}[b]{0.49\textwidth}
         \centering
         \includegraphics[width=0.9\textwidth]{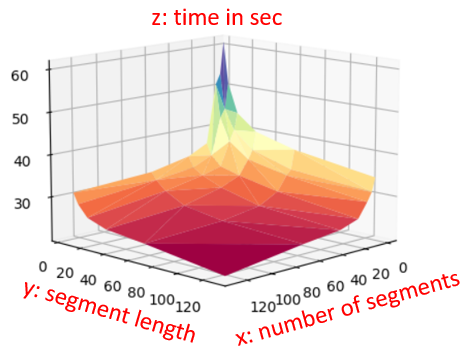}
         \caption{$median$}
         \label{fig:randommedian2}
     \end{subfigure}
        \caption[3D random trace]{3D graphs of average and median run-time in seconds for random trace, with MergeMarathon.}
        \label{fig:3Drandom}
\end{figure}

\begin{figure}[tp!]
     \centering
     \begin{subfigure}[b]{0.49\textwidth}
         \centering
         \includegraphics[width=0.9\textwidth]{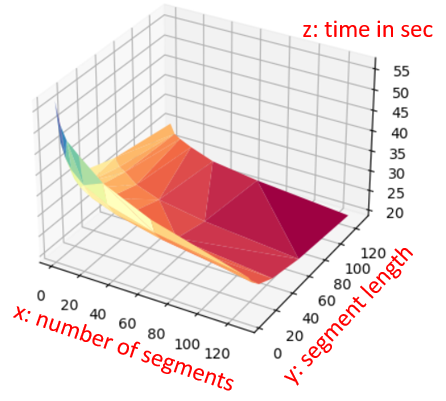}
         \caption{$average$}
         \label{fig:networkaverage1}
     \end{subfigure}
     \hfill
     \begin{subfigure}[b]{0.49\textwidth}
         \centering
         \includegraphics[width=0.9\textwidth]{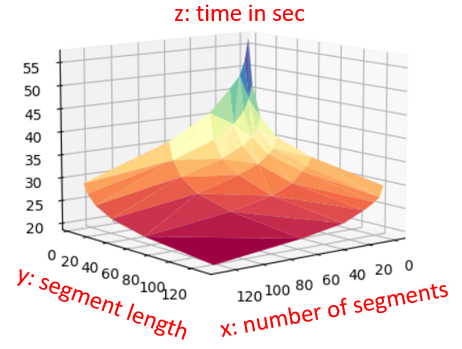}
         \caption{$average$}
         \label{fig:networkaverage2}
     \end{subfigure}
     \hfill
     \begin{subfigure}[b]{0.49\textwidth}
         \centering
         \includegraphics[width=0.9\textwidth]{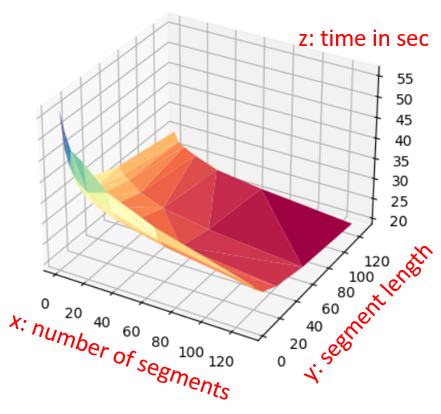}
         \caption{$median$}
         \label{fig:networkmedian1}
     \end{subfigure}
     \hfill
     \begin{subfigure}[b]{0.49\textwidth}
         \centering
         \includegraphics[width=0.9\textwidth]{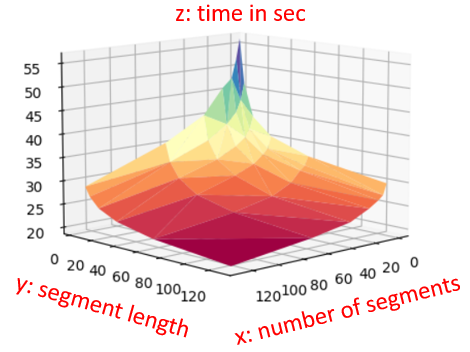}
         \caption{$median$}
         \label{fig:networkmedian2}
     \end{subfigure}
        \caption[3D network trace]{3D graphs of average and median run-time in seconds for network trace, with MergeMarathon.}
        \label{fig:3Dnetwork}
\end{figure}

\begin{figure}[tp!]
     \centering
     \begin{subfigure}[b]{0.49\textwidth}
         \centering
         \includegraphics[width=0.9\textwidth]{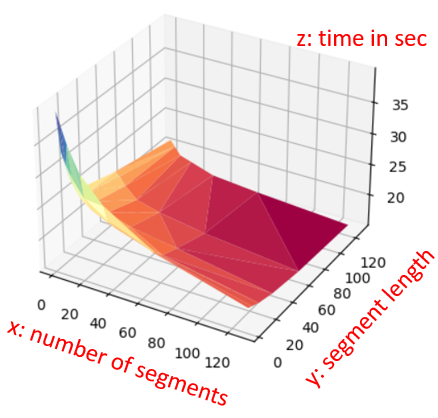}
         \caption{$average$}
         \label{fig:memoryaverage1}
     \end{subfigure}
     \hfill
     \begin{subfigure}[b]{0.49\textwidth}
         \centering
         \includegraphics[width=0.9\textwidth]{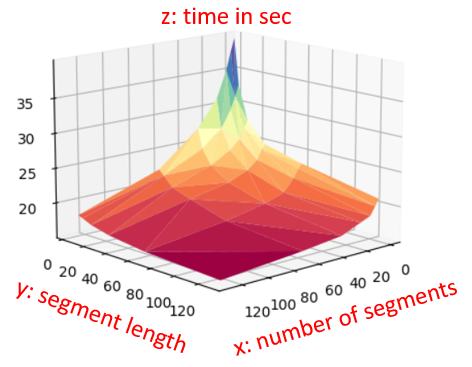}
         \caption{$average$}
         \label{fig:memoryaverage2}
     \end{subfigure}
     \hfill
     \begin{subfigure}[b]{0.49\textwidth}
         \centering
         \includegraphics[width=0.9\textwidth]{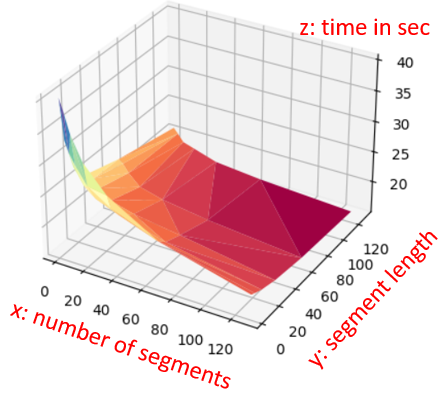}
         \caption{$median$}
         \label{fig:memorymedian1}
     \end{subfigure}
     \hfill
     \begin{subfigure}[b]{0.49\textwidth}
         \centering
         \includegraphics[width=0.9\textwidth]{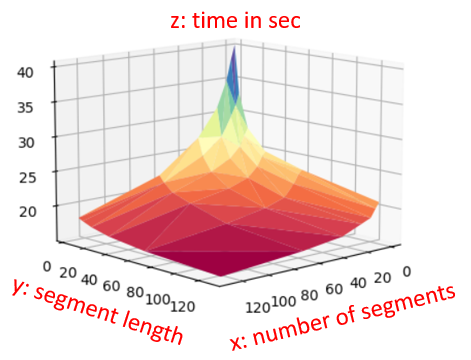}
         \caption{$median$}
         \label{fig:memorymedian2}
     \end{subfigure}
        \caption[3D memory trace]{3D graphs of average and median run-time in seconds for memory trace, with MergeMarathon.}
        \label{fig:3Dmemory}
\end{figure}

We present the results in 3D graphs.
In Figure~\ref{fig:3Drandom} we exhibit the random trace results.
Figure~\ref{fig:3Dnetwork} shows the network trace results and Figure~\ref{fig:3Dmemory} represents the results for the memory trace.
The graphs are similar in their trends, which fit the theoretical analysis.
We can also see that the improvement comes from both the parallelism and the number of stages, and even when only one of them increases and the other one stays fixed, the run-time become shorter. 

\begin{figure}[t]
\begin{center}
\includegraphics[width=0.5\textwidth]{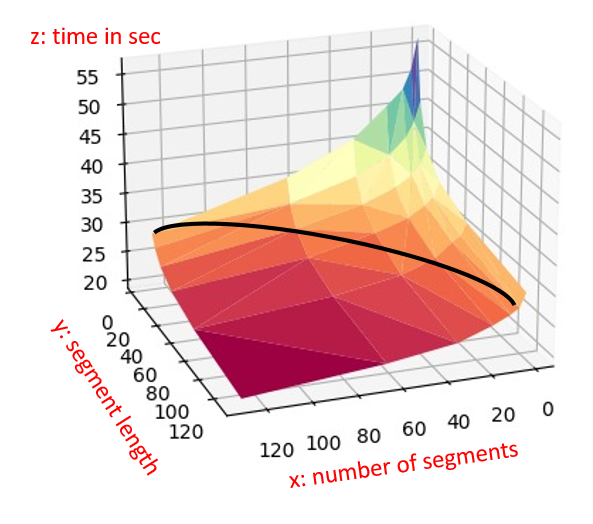}
\end{center}
\caption[3D graph example with line]{3D graph example with line.}
\label{fig:3Dline}
\end{figure}

In Figure~\ref{fig:3Dline}, we can see an example of the random average graph, with a line that separates between the time reduction trends. 
Above the line, when the number of segments and the segments length is under 16, the improvement in the run-time is more significant than under the line, which implies that there are values for number of segments and segments length that satisfy the design trade-off explained in Chapter~\ref{sec:problem}. 
If these values are between the range $(8,16)$, we achieve a significant improvement compared to the original run-time without MergeMarathon. 
Beyond these levels, the returns diminish very quickly.

\subsubsection{Exploring the Performance in 2D}
To understand better the influence of each parameter about the run-time, we present 2D graphs of average and median time. 
There are two types of graphs: 
In the first type, we fixed the segment length, and examined the run-time reduction as a function of the number of segments. 
In the second type, we fixed the number of segments, and examined the run-time reduction as a function of the segment length. 

\begin{figure}[tp!]
     \centering
     \begin{subfigure}[b]{0.49\textwidth}
         \centering
         \includegraphics[width=0.9\textwidth]{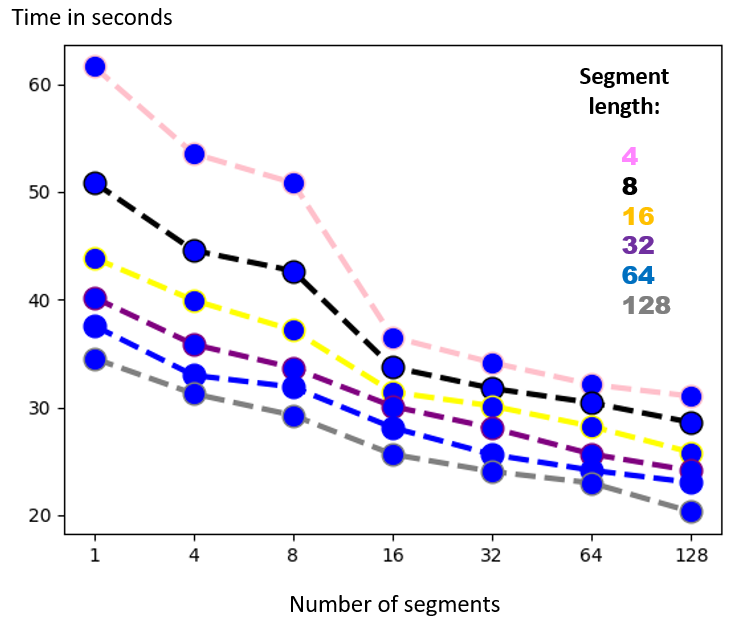}
         \caption{$average$}
         \label{fig:randomaverage1}
     \end{subfigure}
     \hfill
     \begin{subfigure}[b]{0.49\textwidth}
         \centering
         \includegraphics[width=0.9\textwidth]{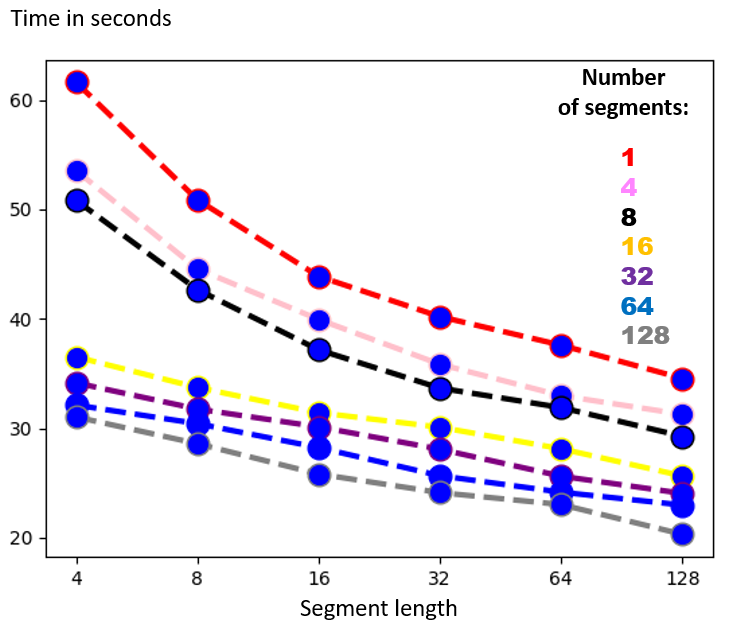}
         \caption{$average$}
         \label{fig:randomaverage2}
     \end{subfigure}
     \hfill
     \begin{subfigure}[b]{0.49\textwidth}
         \centering
         \includegraphics[width=0.9\textwidth]{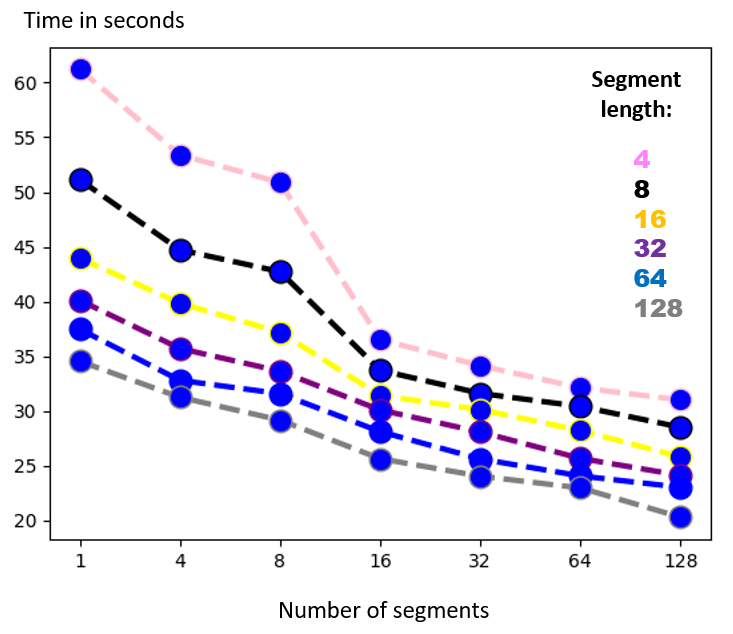}
         \caption{$median$}
         \label{fig:randommedian1}
     \end{subfigure}
     \hfill
     \begin{subfigure}[b]{0.49\textwidth}
         \centering
         \includegraphics[width=0.9\textwidth]{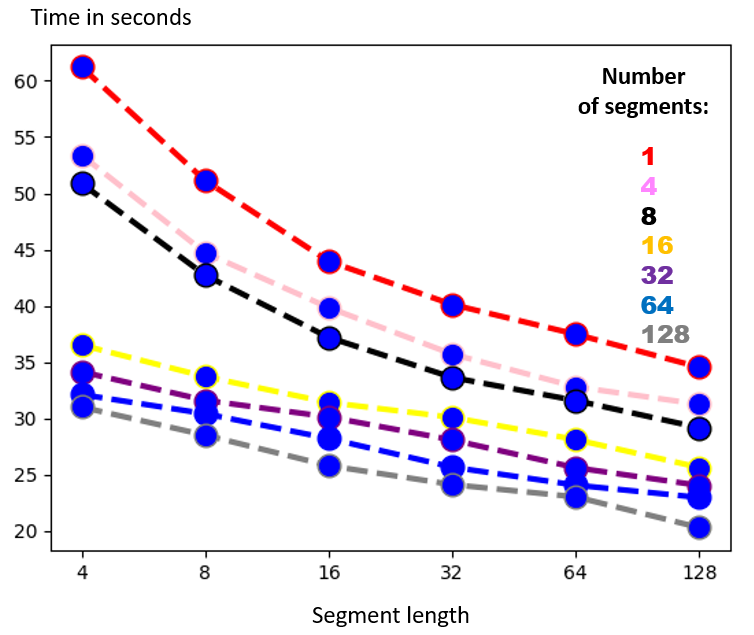}
         \caption{$median$}
         \label{fig:randommedian2}
     \end{subfigure}
        \caption[2D random trace]{2D graphs of average and median run-time in seconds for random trace, with MergeMarathon. Graphs (a) and (c) represent the time reduction as a function of the number of segments, and Graphs (b) and (d) represent the time reduction as a function of the segment length.}
        \label{fig:2Drandom}
\end{figure}

\begin{figure}[tp!]
     \centering
     \begin{subfigure}[b]{0.49\textwidth}
         \centering
         \includegraphics[width=0.9\textwidth]{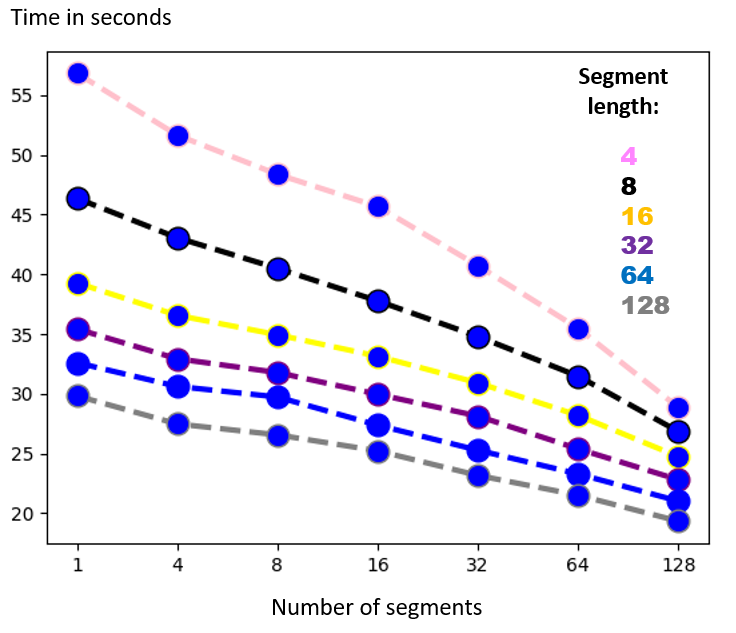}
         \caption{$average$}
         \label{fig:randomaverage1}
     \end{subfigure}
     \hfill
     \begin{subfigure}[b]{0.49\textwidth}
         \centering
         \includegraphics[width=0.9\textwidth]{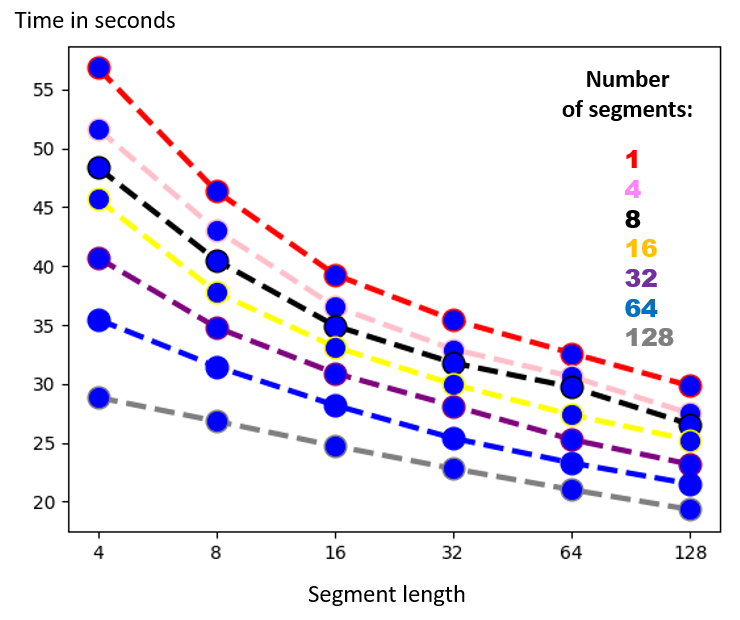}
         \caption{$average$}
         \label{fig:randomaverage2}
     \end{subfigure}
     \hfill
     \begin{subfigure}[b]{0.49\textwidth}
         \centering
         \includegraphics[width=0.9\textwidth]{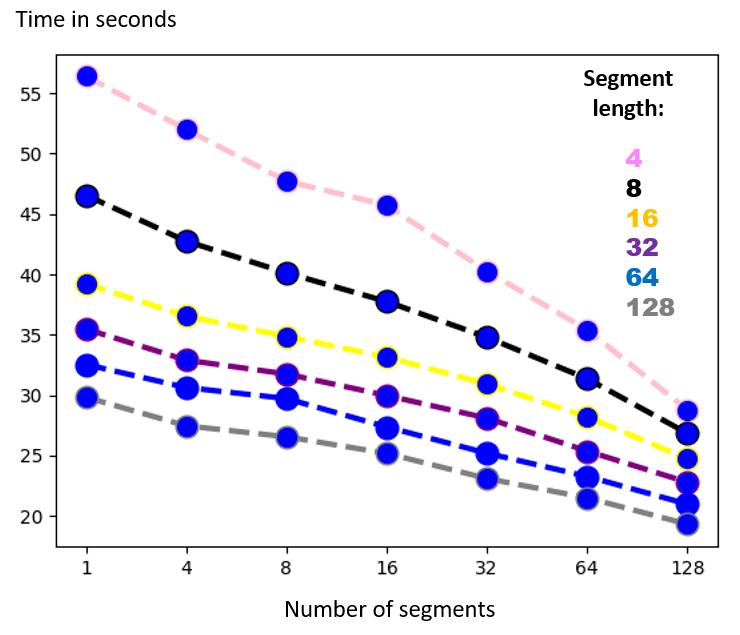}
         \caption{$median$}
         \label{fig:randommedian1}
     \end{subfigure}
     \hfill
     \begin{subfigure}[b]{0.49\textwidth}
         \centering
         \includegraphics[width=0.9\textwidth]{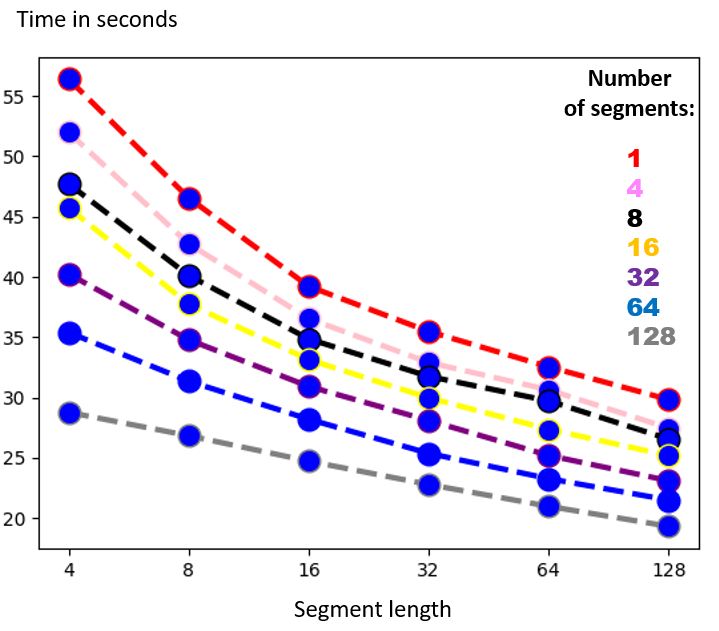}
         \caption{$median$}
         \label{fig:randommedian2}
     \end{subfigure}
        \caption[2D network trace]{2D graphs of average and median run-time in seconds for network trace, with MergeMarathon. Graphs (a) and (c) represent the time reduction as a function of the number of segments, and Graphs (b) and (d) represent the time reduction as a function of the segment length.}
        \label{fig:2Dnetwork}
\end{figure}

\begin{figure}[tp!]
     \centering
     \begin{subfigure}[b]{0.49\textwidth}
         \centering
         \includegraphics[width=0.9\textwidth]{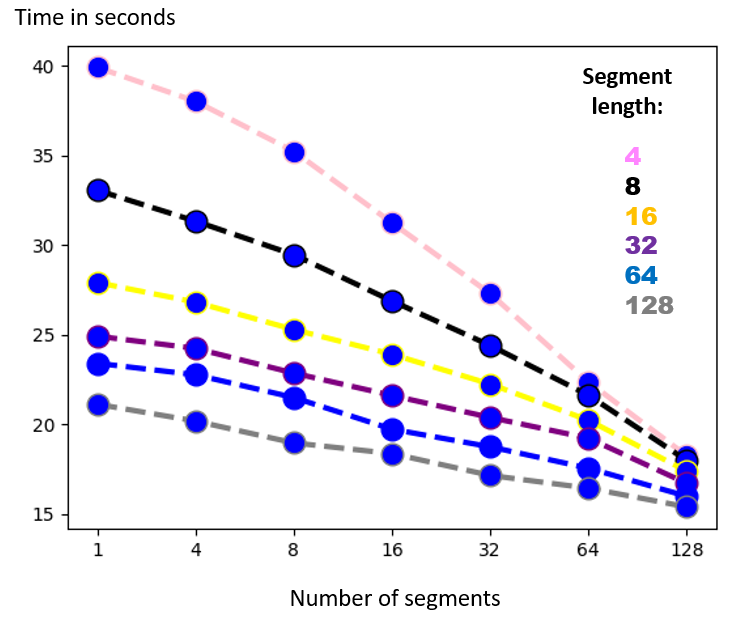}
         \caption{$average$}
         \label{fig:randomaverage1}
     \end{subfigure}
     \hfill
     \begin{subfigure}[b]{0.49\textwidth}
         \centering
         \includegraphics[width=0.9\textwidth]{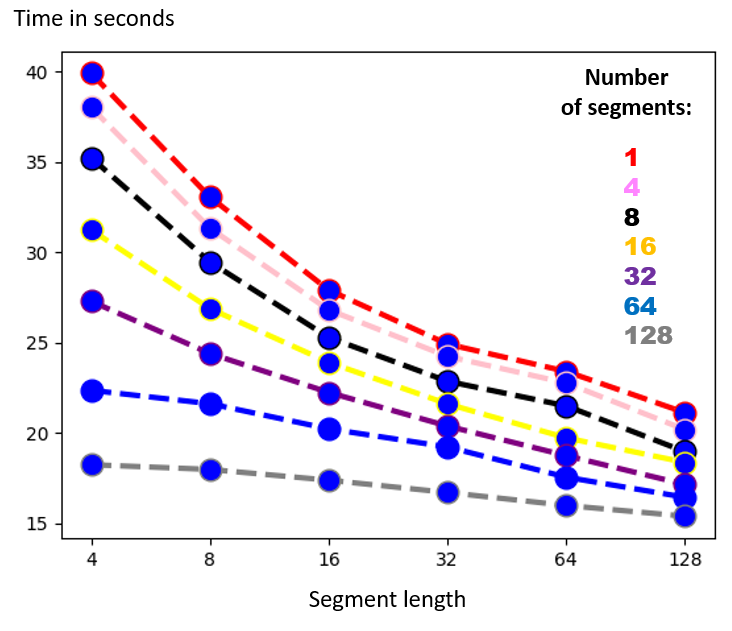}
         \caption{$average$}
         \label{fig:randomaverage2}
     \end{subfigure}
     \hfill
     \begin{subfigure}[b]{0.49\textwidth}
         \centering
         \includegraphics[width=0.9\textwidth]{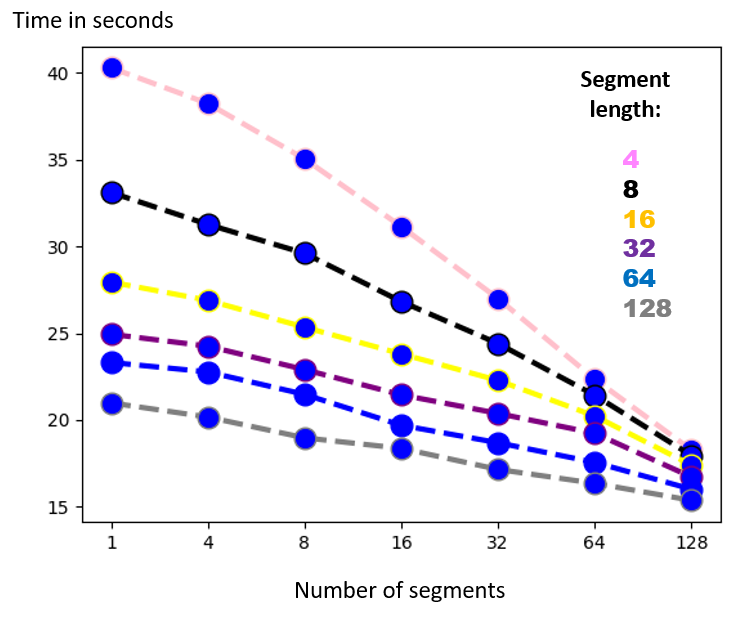}
         \caption{$median$}
         \label{fig:randommedian1}
     \end{subfigure}
     \hfill
     \begin{subfigure}[b]{0.49\textwidth}
         \centering
         \includegraphics[width=0.9\textwidth]{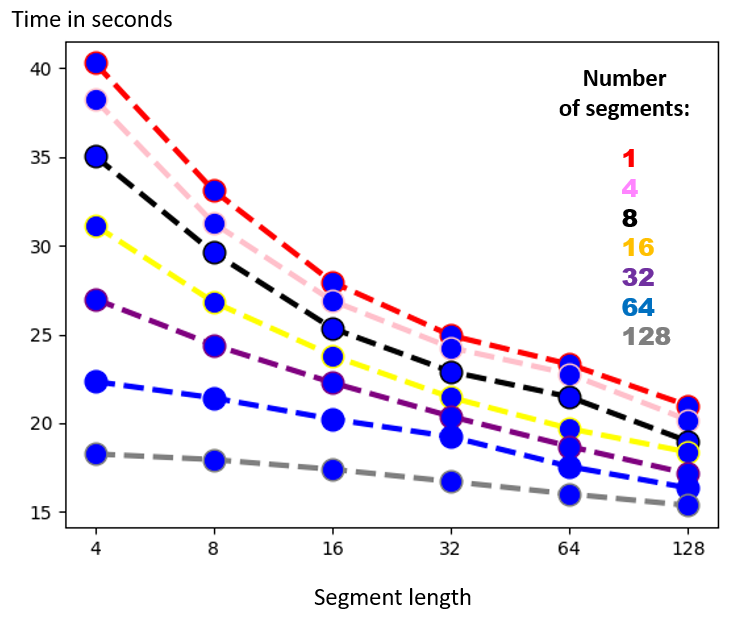}
         \caption{$median$}
         \label{fig:randommedian2}
     \end{subfigure}
        \caption[2D memory trace]{2D graphs of average and median run-time in seconds for memory trace, with MergeMarathon. Graphs (a) and (c) represent the time reduction as a function of the number of segments, and Graphs (b) and (d) represent the time reduction as a function of the segment length.}
        \label{fig:2Dmemory}
\end{figure}

In Figures~\ref{fig:2Drandom},~\ref{fig:2Dnetwork} and~\ref{fig:2Dmemory} we exhibit the 2D graphs of the average and the median run-time in seconds for the random trace with MergeMarathon. 
In each Figure, Graphs (a) and (c) represent the time reduction vs. the number of segments and the different colors represent the different segment lengths. 
Graphs (b) and (d) represent the time reduction as a function of the segment length and the different colors represent the different numbers of segments. 
We analyze the graphs and their trends.

\setcounter{secnumdepth}{3}
\paragraph{Random trace}
In figure~\ref{fig:2Drandom} graphs (a) and (c), we notice that for 4 stages (segment length 4 - the pink line), going from 8 segments to 16 segments leads to a sharp reduction in the run-time, while under 8 segments and above 16 segments the reduction is more moderate. 
We attribute this sharp drop to the fact that at 16 segments, each sub-stream becomes short enough so that the execution of the Merge sort algorithm becomes more cache and virtual memory friendly. 

To validate this assumption, we used valgrind and the Linux time tool on the relevant runs. 
The rate of the cache misses were similar in 8 segments and 16 segments, but the number of the minor page faults dropped down from 3,459,904 for 8 segments to 264,308 with 16 segments. 
The time that the process spends in the system also dropped from 13.78 seconds to 1.74 seconds, which implies that the improvement in time, came both directly from the algorithm itself, and indirectly from system stream lining. 

In the black graph (segment length 8), we also see this sharp reduction between 8 and 16, and in the yellow graph (segment length 16) the reduction is more moderate. 
The other graphs are almost linear, while the x-axis of is logarithmic. 
Thus, for linear x-axis, the graph would be logarithmic, which fits the fact that the server works in parallel on shorter sections and in general the formal analysis.

In graphs (b) and (d), we notice that the red, pink and black graphs (1, 4 and 8 segments) have sharper reduction between 4 and 16 segments, compared to the other graphs (16 segments and above).
We checked the differences between the black graph and the yellow graph in a similar manner to the previous graphs. 
We examined the row of 4 segments of length 4, and found that for 8 segments with length 4 (first dot in the black graph) the number of minor segmentation faults was 3,459,904 and system time is 15.87 seconds, while for 16 segments with length 4 (first dot in the yellow graph) the number of minor segmentation faults was 264,308 and the time spent in the system is 2.06 seconds. 
The cache miss rate was similar in both. 
This also implies that there is indirect improve from the system process.

\setcounter{secnumdepth}{3}
\paragraph{Network trace}
In the network traces the graphs are more moderate. In Figure~\ref{fig:2Dnetwork} graphs (a) and (c), the lines are almost linear (except one dot in the pink graph of 16 segments - likewise the random graphs), which implies that with linear x-axis (in these graphs x-axis is logarithmic), the graphs would be behave like $log(\frac{1}{x})$, which fits the algorithm's analysis, as explained above.
The spaces between the colored graphs are logarithmic, witch fits the theoretical analysis for the segment length.

In graphs (b) and (d) the upper lines show sharper reduction under segment length 16, and moderate reduction above 16. 
This follows the system improvements. 
The lower graphs are more linear, which fits the theoretical analysis for the segment length.

\setcounter{secnumdepth}{3}
\paragraph{Memory trace}
In figure~\ref{fig:2Dmemory}, graphs (a) and (c) shows similar trends to the network traces, and the analysis is similar also. 
The spaces between the graphs point to logarithmic time reduction, which fits to the theoretical analysis for the segment length.
Graphs (b) and (d) are also similar to the network graphs in their trends, spaces and analysis.

\subsection{Summary}
To understand the differences between the traces, we also checked the unique values in each one of them. 
We found that in the random trace there are 32,768 unique values, while the network trace contains only 1,475 and the memory trace even less, only 368 unique values. 
This helps to explain the improvement that comes from the system streamlining in virtual memory, and implies that the run-time improvement is even better when the input is random.
In addition, to verify our theoretical analysis, we checked all the outputs of each configuration of the switch with each one of the traces. 
From each experiment, we collected data from the switch's outputs. 
We collected and analyzed statistics on the Runs, e.g., the average and median length of Runs and the number of Runs. 
The statistics match the theoretical analysis and the actual results, hence verify the results and their analysis.

\section{Discussion}
\label{sec:conclusion}

In our work, we have studied the programmable switches model. 
We explored the model's restrictions and limitations.
We also studied algorithms that accelerate database queries using programmable switches~\cite{tirmazi2020cheetah}. 
These yielded our MergeMarathon algorithm, which accelerates the sorting process at the server side by using programmable switches. 
We created an infrastructure for simulating a programmable switch and a server, and assigned MergeMarathon algorithm to the switch.
We designed MergeMarathon to follow the model and its limitation.

Due to the fact that each switch has its own hardware properties, such as parallelism level and number of stages in the pipeline, our simulated switch infrastructure enables to set these numbers. 
Hence, we examined our algorithm on various such parameters. 
Our experiments showed that MergeMarathon improved the run-time at the server side, even in switches with only one pipeline segment, i.e., with no parallelism at all. 
When we used parallelism, there was also a significant improvement, that is increased with the number of the stages in the pipeline.
In any case, MergeMarathon reduced the run-time at the server side by at least $20\%$ and up to $75\%$ in the best cases.
The average improvement in our tests is $50\%$ of the run-time.

\subsubsection*{Future Work}
Looking into the future, we would like to extend our MergeMarathon to be implemented in the P4 language, and applied to a real programmable switches, that placed inside the network. 
We would like to analyze the savings in run-time and resources at the server with real database queries sent to the server.

\bibliographystyle{abbrv}
\bibliography{references}
\end{document}